\documentclass[english,aps,nofootinbib,twocolumn,eqsecnum,reprint,superscriptaddress]{revtex4-1}
\usepackage[T1]{fontenc}
\usepackage[utf8]{luainputenc}
\usepackage{color}
\usepackage{amsmath}
\usepackage{amssymb}
\usepackage{graphicx}
\usepackage{setspace}
\usepackage{esint}

\makeatletter

\usepackage[T1]{fontenc}
\usepackage[utf8]{luainputenc}

\usepackage{array}
\usepackage{graphicx,amsmath,amsfonts,amssymb,slashed}
\usepackage{appendix}

\allowdisplaybreaks[4]

\usepackage{float}
\usepackage{wasysym}

\usepackage{esint}
\usepackage[marginal]{footmisc}
\usepackage{braket}
\usepackage{numprint}
\usepackage{enumitem}

\usepackage{verbatim}
\usepackage{numprint}
\usepackage{lipsum}
\usepackage{enumitem}
\usepackage{amsfonts}
\usepackage{mathrsfs}
\usepackage{scrextend}

\usepackage{setspace}

\renewcommand{\baselinestretch}{1}
\makeatother

\makeatother

\usepackage{babel}
\begin{document}
\title{In-medium screening effects for the Galactic halo and solar-reflected
\\
dark matter detection in semiconductor targets}
\author{Zheng-Liang Liang}
\email{liangzl@mail.buct.edu.cn}

\affiliation{College of Mathematics and Physics, Beijing University of Chemical
Technology~\\
Beijing 100029, China}
\author{Chongjie Mo}
\email{cjmo@csrc.ac.cn}

\affiliation{Beijing Computational Science Research Center,~\\
Beijing, 100193, China}
\author{Ping Zhang}
\email{pzhang2012@qq.com}

\affiliation{School of Physics and Physical Engineering, Qufu Normal University~\\
 Qufu, 273165, China}
\affiliation{Institute of Applied Physics and Computational Mathematics~\\
Beijing, 100088, China}
\begin{abstract}
Recently, the importance of the electronic many-body effect in the
dark matter~(DM) detection has been recognized and a coherent formulation
of the DM-electron scattering in terms of the dielectric response
of the target material has been well established in literatures. In
this paper, we put relevant formulas into practical density functional
theory~(DFT) estimation of the excitation event rates for the diamond
and silicon semiconductor targets. Moreover, we compare the event
rates calculated from the energy loss functions with and without the
local field effects. For a consistency check of this numerical method,
we also compare the differential spectrum and detection reach of the
silicon target with those computed with the $\mathtt{GPAW}$ code.
It turns out that these DFT approaches are quite consistent and robust.
As an interesting extension, we also investigate the in-medium effect
on the detection of the solar-reflected DM flux in silicon-based detectors,
where the screening effect is found to be less remarkable than on
detection of the Galactic DM, due to the high energies of the reflected
DM particles. 
\end{abstract}
\maketitle

\section{Introduction}

In recent years, both theorists and experimentalists begin to shift
their focus on other directions beyond the weakly interacting massive
particles~(WIMPs). The sub-GeV dark matter~(DM) as an alternative
candidate, has attracted increasing attention for its theoretical
motivations and detection feasibility. In the sub-GeV DM paradigm,
the DM particles are expected to reveal itself via the weak DM-electron
interaction in silicon- and germanium-based semiconductors~(e.g.,
SENSEI~\citep{Barak:2020fql}, DAMIC~\citep{Castello-Mor:2020jhd},
SuperCDMS~\citep{Amaral:2020ryn}, and EDELWEISS~\citep{Arnaud:2020svb})
with energy thresholds as low as a few eV. In the theoretical aspect,
since the appearance of the first estimation of the electronic excitation
rates based on the first-principles density functional theory~(DFT)~\citep{Essig:2015cda},
similar investigations have been generalized to a wider range of target
materials~\citep{Hochberg:2015pha,Essig:2016crl,Hochberg:2016ntt,Hochberg:2016sqx,Derenzo:2016fse,Hochberg:2017wce,Kadribasic:2017obi,Budnik:2017sbu,Knapen2018,Griffin:2018bjn,Griffin:2019mvc,Trickle:2019ovy,Kurinsky:2019pgb,Trickle:2019nya,Campbell-Deem:2019hdx,Coskuner:2019odd,Geilhufe:2019ndy,Griffin:2020lgd},
and have spurred further discussions on the methodology~\citep{Liang:2018bdb,Kurinsky:2020dpb,Kozaczuk:2020uzb,Griffin:2021znd,Hochberg:2021pkt,Knapen:2021run},
and extensive interpretations of the DM-electron interactions~\citep{Baxter:2019pnz,Emken:2019tni,Essig:2019xkx,Heikinheimo:2019lwg,Trickle:2020oki,Andersson:2020uwc,Su:2020zny,Catena:2021qsr}.

Recently, nontrivial collective behavior of the electrons in solid
detectors has also attracted attention~\citep{Kurinsky:2020dpb,Gelmini_2020,Kozaczuk:2020uzb,Knapen:2020aky,Liang:2020ryg,Mitridate:2021ctr}.
The related physics such as screening and the plasmon excitation that
cannot be explained in terms of standard two-body scattering, and
non-interacting single-particle states, can be well described with
the dielectric function. The in-medium effect induced by the DM-electron
interaction has been thoroughly investigated in Refs.~\citep{Knapen:2021run,Hochberg:2021pkt}.
In this work we also touch on this topic. Our first purpose is to
provide a detailed derivation of the the DM-electron excitation event
rate in the context of the linear response theory, and then calculate
the excitation event rates for diamond and silicon targets using the
DFT approach. We begin with the well-established description of the
electron energy loss spectroscopy~(EELS) in the homogeneous electron
gas~(HEG), and generalize the description to the crystalline environments,
and finally to the case of DM-electron excitation process in semiconductor
targets.

As is well known, the key quantity in describing the in-medium effect
in EELS and DM-electron excitation process is the energy loss function~(ELF),
which is defined as the imaginary part of the inverse dielectric function
$\mathrm{Im}\left[-1/\epsilon\left(\mathbf{Q},\omega\right)\right]$
for the HEG, with $\mathbf{Q}$ being the momentum and $\omega$ being
the energy transferred to the electrons from the impinging particle.
However, for the crystal targets, the ELF is generalized accordingly
to the matrix form $\mathrm{Im}\left[\epsilon_{\mathbf{G},\mathbf{G}'}^{-1}\left(\mathbf{q},\omega\right)\right]$,
where $\mathbf{G}$ and $\mathbf{G}'$ are reciprocal lattice vectors,
and $\mathbf{q}$, as the remainder part of the momentum transfer
$\mathbf{Q}$, is the uniquely determined in the first Brillouin zone~(1BZ).
As will be seen from the following discussions, only the diagonal
components of the inverse dielectric function are relevant for the
description of the screening effect, if the crystal structure is approximated
as isotropic. In this case, the effective inverse dielectric function
$\mathrm{Im}\left[-1/\epsilon\left(\mathbf{Q},\omega\right)\right]$
is approximated as the diagonal components $\mathrm{Im}\left[\epsilon_{\mathbf{G},\mathbf{G}}^{-1}\left(\mathbf{q},\omega\right)\right]$
averaged over $\mathbf{G}$ and $\mathbf{q}$. This treatment includes
the so-called local field effects~(LFEs), as the information of the
off-diagonal components enters the inverse dielectric function.

As mentioned in Ref.~\citep{Knapen:2021run}, there exists an alternative
definition of ELF, where one first averages the diagonal elements
$\epsilon_{\mathbf{G},\mathbf{G}}\left(\mathbf{q},\omega\right)$
over $\mathbf{G}$ and $\mathbf{q}$ to obtained an effective dielectric
function $\overline{\epsilon}\left(\mathbf{Q},\omega\right)$, and
then the inverse dielectric function is approximated as $\mathrm{Im}\left[-1/\overline{\epsilon}\left(\mathbf{Q},\omega\right)\right]$.
In this case, the LFEs are not included. Thus, another purpose of
this work is to give a quantitative comparison between the event rates
obtained from these two inverse dielectric functions. i.e., to investigate
the implication of the LFEs. In addition, we also compare the $\mathtt{YAMBO}$
estimation of the sensitivities of silicon detector with those calculated
using the $\mathtt{GPAW}$ package~\citep{Knapen:2021run}. Although
the ELF has been well formulated and calculated in Ref.~\citep{Knapen:2021run},
it is interesting to perform a consistency check on different numerical
approaches.

As an interesting generalization of above discussion, we also investigate
the screening effect in semiconductor detectors in response to the
solar reflection of leptophilc DM particles. While the conventional
detection strategies are sensitive only to the DM mass above the MeV
scale, probing the solar-reflected DM particles offers new possibility
of extending detection reach down to mass range below the MeV scale~\citep{An:2017ojc,Chen:2020gcl,Emken:2021lgc}.
In this scenario, the hot solar electron gas has a chance to boost
the passing-by halo DM particles to a speed much higher than the galactic
escape velocity, and consequently a sub-MeV DM particle is able to
trigger ionization signals in conventional detectors. Unlike the case
of the halo DM where excitation event spectra fall off quickly in
energy region above a few tens of eV, the event spectra of the solar
reflection extend far into higher energy range, which may bring different
features of screening effect in detecting the solar-reflected DM flux.

This paper is organized as follows. In Sec.~\ref{sec:EELS} we first
take a review of the EELS in both electron gas and crystalline structure,
respectively. Based on these discussions, we then further derive relevant
formulas for the excitation rate induced by the DM-electron scattering.
In Sec.~\ref{sec:Solar_Reflection}, we first calculate the solar-reflected
DM flux from Monte Carlo simulation approach, and then investigate
the in-medium effect in detection of reflected DM signals. We conclude
in Sec.~\ref{sec:conclusions}.

\section{\label{sec:EELS}From EELS to DM-induced excitation}

In this section, we take a brief review of theoretical description
of the EELS in HEG and in crystalline solids, and extend the formalism
to include the electronic excitation process induced by the incident
DM particle, in the context of the ELFs.

\subsection{EELS in electron gas }

\begin{figure}[h]
\begin{centering}
\includegraphics[scale=0.55]{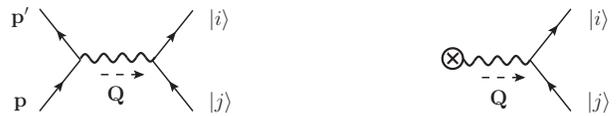} 
\par\end{centering}
\caption{\textit{\label{fig:EELS}}\textbf{\textit{Left}}: The impinging electron
collides with another electron in the target material, and excites
the latter from state $\ket{j}$ to state $\ket{i}$. \textbf{\textit{Right}}:
An equivalent description where the effect of incident electron is
represented with a source term. See text for details.}
\end{figure}

The EELS provides the spectrum information of the energy transferred
from a fast impinging electron to the target material, which is deposited
either in the form of electron-hole pairs, or collective excitations~(plasmons).
We begin the discussion with the diagram in the left panel in Fig.~\ref{fig:EELS}
that describes the process where one incident electron excites another
in the target material from state $\ket{j}$ to state $\ket{i}$.
With the Feynman rules summarized in appendix of Ref.~\citep{Liang:2020ryg},
relevant amplitude reads as 
\begin{eqnarray}
i\mathcal{M} & = & -iV_{\mathrm{Cou}}\left(\mathbf{Q}\right)\braket{i|e^{i\mathbf{Q}\cdot\hat{\mathbf{x}}}|j}\nonumber \\
 & = & -i\frac{4\pi\alpha}{Q^{2}}\braket{i|e^{i\mathbf{Q}\cdot\hat{\mathbf{x}}}|j},
\end{eqnarray}
where $\mathbf{Q}=\mathbf{p}-\mathbf{p}'$, with $\mathbf{p}$ ($\mathbf{p}'$)
is the electron momentum before (after) the scattering, $V_{\mathrm{Cou}}$
represents the propagator of the electron-electron Coulomb interaction,
and $\alpha$ is the electromagnetic fine structure constant. To calculate
the cross section, one needs to average over the initial states and
sum over the final states of electrons in crystal, at a finite temperature
$T$, so it is more convenient to treat this problem in the context
of the linear response theory. To this end, the effects brought by
the incident electron is regarded as a perturbation exerted onto the
electronic system of the target material, which can be summarized
as the following effective Hamiltonian for the electrons in solids~(i.e.,
the source term illustrated in the right panel in Fig.~\ref{fig:EELS}):
\begin{align}
 & \hat{H}_{I}\left(t\right)=\hat{H}_{I}e^{i\omega_{p'p}t}\nonumber \\
 & =V_{\mathrm{Cou}}\left(\mathbf{Q}\right)\int e^{i\mathrm{\mathbf{Q}\cdot\mathbf{x}}}\,\hat{\rho}_{I}\left(\mathbf{x},t\right)\mathrm{d}^{3}x\,e^{i\omega_{p'p}t}\nonumber \\
 & =V_{\mathrm{Cou}}\left(\mathbf{Q}\right)\int e^{i\mathrm{\mathbf{Q}\cdot\mathbf{x}}}\,\hat{\psi}_{I}^{\dagger}\left(\mathbf{x},t\right)\hat{\psi}_{I}\left(\mathbf{x},t\right)\mathrm{d}^{3}x\,e^{i\omega_{p'p}t},\label{eq:density-density}
\end{align}
where $\hat{\rho}_{I}$ and $\hat{\psi}_{I}$ are the density and
the field operators of the electron, respectively, and  $\omega_{p'p}=p'^{2}/2m_{e}-p^{2}/2m_{e}$
is the energy difference between the outgoing and incoming electron~($m_{e}$
is the electron mass). Thus the averaging and summing procedure can
be expressed as a correlation function 
\begin{eqnarray}
S_{\hat{H}_{I}^{\dagger}\hat{H}_{I}}\left(-\omega_{p'p}\right) & = & \sum_{i,\,j}p_{j}\left|\braket{i|\hat{H}_{I}|j}\right|^{2}\left(2\pi\right)\delta\left(\varepsilon_{i}-\varepsilon_{j}+\omega_{p'p}\right)\nonumber \\
 & = & \int_{-\infty}^{+\infty}\left\langle \hat{H}_{I}^{\dagger}\left(0\right)\hat{H}_{I}\left(t\right)\right\rangle e^{i\omega_{p'p}t}\mathrm{d}t,
\end{eqnarray}
where $p_{j}$ is the thermal distribution of the initial state $\ket{j}$,
and the symbol $\left\langle \cdots\right\rangle $ represents the
thermal average. At this stage, one can insert this correlation function
into the formula for the cross section~(Fermi's golden rule) in terms
of the inverse dielectric function $\epsilon^{-1}\left(\mathbf{Q},\omega\right)$,
\begin{eqnarray}
\sigma & = & \int\frac{\mathrm{d}^{3}Q\,\mathrm{d}^{3}p'}{\left(2\pi\right)^{3}}\frac{S_{\hat{H}_{I}^{\dagger}\hat{H}_{I}}\left(\omega\right)\delta^{3}\left(\mathbf{p}'-\mathbf{p}+\mathbf{Q}\right)\delta\left(\omega_{p'p}+\omega\right)}{v}\mathrm{d}\omega\nonumber \\
 & \simeq & V\int\frac{\mathrm{d}^{3}Q}{\left(2\pi\right)^{3}}\frac{2V_{\mathrm{Cou}}\left(\mathbf{Q}\right)}{v}\mathrm{Im}\left[\frac{-1}{\epsilon\left(\mathbf{\mathbf{Q}},\omega\right)}\right]\nonumber \\
 &  & \times\delta\left(\frac{\left|\mathbf{Q}\right|^{2}}{2m_{e}}-\mathbf{v}\cdot\mathbf{Q}+\omega\right)\mathrm{d}\omega,\label{eq:HEG_cross_section}
\end{eqnarray}
where $v$ is the velocity of the incident electron, and $V$ represents
the volume of the material. In above derivation we utilize the fluctuation-dissipation
theorem
\begin{align}
 & S_{\hat{H}_{I}^{\dagger}\hat{H}_{I}}\left(\omega\right)=i\frac{\left[\chi_{\hat{H}_{I}^{\dagger}\hat{H}_{I}}\left(\omega+i0^{+}\right)-\chi_{\hat{H}_{I}^{\dagger}\hat{H}_{I}}\left(\omega-i0^{+}\right)\right]}{1-e^{-\beta\omega}}\nonumber \\
 & \simeq iV\left|V_{\mathrm{Cou}}\left(\mathbf{Q}\right)\right|^{2}\,\left[\chi_{\hat{\rho}\hat{\rho}}\left(\mathbf{Q},\,\omega+i0^{+}\right)-\chi_{\hat{\rho}\hat{\rho}}\left(\mathbf{Q},\,\omega-i0^{+}\right)\right]\nonumber \\
 & =-2V\left|V_{\mathrm{Cou}}\left(\mathbf{Q}\right)\right|^{2}\,\mathrm{Im}\left[\chi_{\hat{\rho}\hat{\rho}}^{\mathrm{r}}\left(\mathbf{\mathbf{Q}},\,\omega\right)\right]\nonumber \\
 & =2V\,V_{\mathrm{Cou}}\left(\mathbf{Q}\right)\,\mathrm{Im}\left[\frac{-1}{\epsilon\left(\mathbf{\mathbf{Q}},\omega\right)}\right],\label{eq:Shh}
\end{align}
where $\beta=1/T$ is the inverse temperature, and we adopt the zero-temperature
approximation $1-e^{-\beta\omega}\approx1$; $\chi_{\hat{A}\hat{B}}\left(z\right)$
is the master function of the correlation functions of the operators
$\hat{A}$ and $\hat{B}$, which yields relevant retarded correlation
function $\chi_{\hat{A}\hat{B}}^{\mathrm{r}}\left(\omega\right)=\chi_{\hat{A}\hat{B}}\left(\omega+i0^{+}\right)$
and advanced correlation function $\chi_{\hat{A}\hat{B}}^{\mathrm{a}}\left(\omega\right)=\chi_{\hat{A}\hat{B}}\left(\omega-i0^{+}\right)$
in momentum space; the inverse dielectric function in the last line
connects the retarded density-density correlation function $\chi_{\hat{\rho}\hat{\rho}}^{\mathrm{r}}\left(\mathbf{\mathbf{Q}},\,\omega\right)$
through the following relation,
\begin{eqnarray}
\frac{1}{\epsilon\left(\mathbf{Q},\omega\right)} & = & 1+V_{\mathrm{Cou}}\left(\mathbf{Q}\right)\chi_{\hat{\rho}\hat{\rho}}^{\mathrm{r}}\left(\mathbf{Q},\,\omega\right).\label{eq:inverse_epsilon0}
\end{eqnarray}
On the other hand, the Schwinger-Dyson equation for the screen Coulomb
interaction connects the dielectric function and the the polarizability
$\Pi\left(\mathbf{Q},\,\omega\right)$ through the relation 
\begin{eqnarray}
\epsilon\left(\mathbf{Q},\,\omega\right) & = & 1-V_{\mathrm{Cou}}\left(\mathbf{Q}\right)\Pi\left(\mathbf{Q},\,\omega\right).
\end{eqnarray}
In the random phase approximation~(RPA), $\Pi\left(\mathbf{Q},\,\omega\right)$
is approximated by the electron-hole loop and thus the dielectric
function can be expressed as
\begin{eqnarray}
\epsilon\left(\mathbf{\mathbf{Q}},\omega\right) & \simeq & 1-\frac{V_{\mathrm{Cou}}\left(\mathbf{Q}\right)}{V}\sum_{i,j}\frac{\left|\braket{i|e^{i\mathbf{\mathbf{Q}}\cdot\hat{\mathbf{x}}}|j}\right|^{2}}{\varepsilon_{i}-\varepsilon_{j}-\omega-i0^{+}}\left(n_{i}-n_{j}\right),\nonumber \\
\label{eq:RPA-dielectric-fucnition0}
\end{eqnarray}
where $n_{i}$~($n_{j}$) and $\varepsilon_{i}$~($\varepsilon_{j}$)
denote the occupation number and the energy of the state $\ket{i}$~($\ket{j}$),
respectively. Plugging the dielectric function Eq.~(\ref{eq:RPA-dielectric-fucnition0})
into Eq.~(\ref{eq:HEG_cross_section}) yields the EELS cross section
for the HEG.

\subsection{EELS in crystalline solids}

Above discussion of the EELS for the HEG can be straightforwardly
extended to the case in crystal structure, as long as one takes into
consideration the LFEs in the crystalline environment. In crystalline
solid where the translational symmetry for continuous space reduces
to that for the crystal lattice, the correlation functions can no
longer be expressed as differences of the space-time coordinates.
In this case, any function periodic in position $\chi\left(\mathbf{x},\mathbf{x}';\omega\right)$
can be expressed in the reciprocal space as the following, 
\begin{eqnarray}
\chi\left(\mathbf{x},\mathbf{x}';\omega\right) & = & \frac{1}{V}\sum_{\mathbf{k}\in1\mathrm{BZ}}\sum_{\mathbf{G},\mathbf{G}'}e^{i\left(\mathbf{k}+\mathbf{G}\right)\cdot\mathbf{x}}\nonumber \\
 &  & \times\chi_{\mathbf{G},\mathbf{G}'}\left(\mathbf{k},\omega\right)e^{-i\left(\mathbf{k}+\mathbf{G}'\right)\cdot\mathbf{x}'}
\end{eqnarray}
where $\chi_{\mathbf{G},\mathbf{G}'}\left(\mathbf{k};\omega\right)$
is the reciprocal matrix with $\mathbf{G}$ and $\mathbf{G}'$ being
reciprocal lattice vectors and $\mathbf{k}$ is restricted to the
1BZ, which can be determined with the Fourier transformation 
\begin{eqnarray}
\chi_{\mathbf{G},\mathbf{G}'}\left(\mathbf{k},\omega\right) & = & \frac{1}{V}\int\mathrm{d}^{3}x\,\mathrm{d}^{3}x\,'e^{-i\left(\mathbf{k}+\mathbf{G}\right)\cdot\mathbf{x}}\nonumber \\
 &  & \times\chi\left(\mathbf{x},\mathbf{x}';\omega\right)\,e^{i\left(\mathbf{k}+\mathbf{G}'\right)\cdot\mathbf{x}'}.\label{eq:inverse GG}
\end{eqnarray}
As a consequence, for an arbitrary momentum transfer $\mathbf{Q}$,
which can be split into a reduced momentum confined in the 1BZ, and
a reciprocal one, i.e., $\mathbf{\mathbf{Q}}=\mathbf{q}+\mathbf{G}$,
one obtains the following correspondence in crystalline environment:
\begin{eqnarray}
\chi_{\hat{\rho}\hat{\rho}}^{\mathrm{r}}\left(\mathbf{Q},\,\omega\right) & \rightarrow & \chi_{\hat{\rho}\hat{\rho}\,\mathbf{G},\mathbf{G}}^{\mathrm{r}}\left(\mathbf{q},\omega\right)\nonumber \\
 & = & \frac{1}{V}\int\mathrm{d}^{3}x\,\mathrm{d}^{3}x\,'e^{-i\left(\mathbf{q}+\mathbf{G}\right)\cdot\mathbf{x}}\nonumber \\
 &  & \times\chi_{\hat{\rho}\hat{\rho}}^{\mathrm{r}}\left(\mathbf{x},\mathbf{x}';\omega\right)\,e^{i\left(\mathbf{q}+\mathbf{G}\right)\cdot\mathbf{x}'}.
\end{eqnarray}
$\chi_{\hat{\rho}\hat{\rho}\,\mathbf{G},\mathbf{G}'}^{\mathrm{r}}$
is connected to the inverse microscopic dielectric matrix $\epsilon_{\mathbf{G},\mathbf{G}'}$
through the relation 
\begin{widetext}
\begin{eqnarray}
\epsilon_{\mathbf{G},\mathbf{G}'}^{-1}\left(\mathbf{q},\omega\right) & = & \delta_{\mathbf{G},\mathbf{G}'}+V_{\mathbf{G},\mathbf{G}'}^{\mathrm{Cou}}\left(\mathbf{q}\right)\chi_{\hat{\rho}\hat{\rho}\,\mathbf{G},\mathbf{G}'}^{\mathrm{r}}\left(\mathbf{q},\,\omega\right),
\end{eqnarray}
where $V_{\mathbf{G},\mathbf{G}'}^{\mathrm{Cou}}\left(\mathbf{q}\right)$=$V_{\mathrm{Cou}}\left(\mathbf{q}+\mathbf{G}\right)\delta_{\mathbf{G},\mathbf{G}'}$=$4\pi\alpha\delta_{\mathbf{G},\mathbf{G}'}/\left|\mathbf{q}+\mathbf{G}\right|^{2}$
is obtained from Eq.~(\ref{eq:inverse GG}). Consequently, the expression
for the cross section for the HEG in Eq.~(\ref{eq:HEG_cross_section})
can be extended to the case in crystal structure as follows, 

\begin{eqnarray}
\sigma & \simeq & V\sum_{\mathbf{G}}\int_{1\mathrm{BZ}}\frac{\mathrm{d}^{3}q}{\left(2\pi\right)^{3}}\frac{2\,V_{\mathrm{Cou}}\left(\mathbf{G}+\mathbf{q}\right)}{v}\,\mathrm{Im}\left[\epsilon_{\mathbf{G},\mathbf{G}}^{-1}\left(\mathbf{q},\omega\right)\right]\,\delta\left[\frac{\left|\mathbf{G}+\mathbf{q}\right|^{2}}{2m_{e}}-\mathbf{v}\cdot\mathbf{\left(\mathbf{G}+\mathbf{q}\right)}+\omega\right]\mathrm{d}\omega\nonumber \\
 & = & \sum_{\mathbf{G}}\sum_{\mathbf{\mathbf{q}\in1\mathrm{BZ}}}\int\frac{2\,V_{\mathrm{Cou}}\left(\mathbf{G}+\mathbf{q}\right)}{v}\,\mathrm{Im}\left[\epsilon_{\mathbf{G},\mathbf{G}}^{-1}\left(\mathbf{q},\omega\right)\right]\,\delta\left[\frac{\left|\mathbf{G}+\mathbf{q}\right|^{2}}{2m_{e}}-\mathbf{v}\cdot\mathbf{\left(\mathbf{G}+\mathbf{q}\right)}+\omega\right]\mathrm{d}\omega.\label{eq:cross_sectioon_crystal}
\end{eqnarray}
In this study, we adopt the RPA for the microscopic dielectric matrix:
\begin{eqnarray}
\epsilon_{\mathbf{G},\mathbf{G}'}\left(\mathbf{q},\omega\right) & = & \delta_{\mathbf{G},\mathbf{G}'}-\frac{V_{\mathbf{G},\mathbf{G}}^{\mathrm{Cou}}\left(\mathbf{q}\right)}{V}\sum_{i,j}\frac{\braket{i|e^{i\left(\mathbf{\mathbf{\mathbf{q}+\mathbf{G}}}'\right)\cdot\hat{\mathbf{x}}}|j}\braket{j|e^{-i\left(\mathbf{\mathbf{\mathbf{q}+\mathbf{G}}}\right)\cdot\hat{\mathbf{x}}}|i}}{\varepsilon_{ij}-\omega-i0^{+}}\left(n_{i}-n_{j}\right).\label{eq:RPA matrix}
\end{eqnarray}
In practice, the inverse dielectric function $\epsilon_{\mathbf{G},\mathbf{G}}^{-1}\left(\mathbf{q},\omega\right)$
is obtained via directly inverting the dielectric matrix in Eq.~(\ref{eq:RPA matrix}).
\end{widetext}

\subsection{\label{sec:Electron-DM}Electron excitation induced by DM particles}

\begin{figure}[h]
\begin{centering}
\includegraphics[scale=0.55]{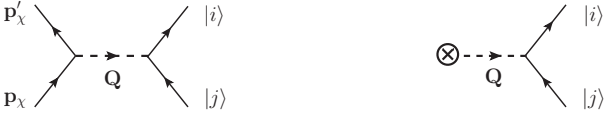} 
\par\end{centering}
\caption{\label{fig:DM-Electron}\textbf{\textit{Left}}: Feynman diagram for
the impinging DM particle collides with an electron in the target
material, and excites it from state $\ket{j}$ to state $\ket{i}$.
\textbf{\textit{Right}}: An equivalent description where the effect
of incident DM particle is replaced with a source term.}
\end{figure}

Above discussion of the EELS can be transplanted in a parallel manner
to the scenario where the impinging particle is a DM particle~(see
Fig.~\ref{fig:DM-Electron}). In this case, the Coulomb potential
$V_{\mathrm{Cou}}\left(\mathbf{Q}\right)$ should be replaced by the
DM-electron interaction $V_{\chi e}\left(\mathbf{Q}\right)$ leading
to Eq.~(\ref{eq:Shh}), which yields 
\begin{eqnarray}
S_{\hat{H}_{I}^{\dagger}\hat{H}_{I}}\left(\omega\right) & = & 2V\,\frac{\left|V_{\chi e}\left(\mathbf{Q}\right)\right|^{2}}{V_{\mathrm{Cou}}\left(\mathbf{\mathbf{Q}}\right)}\mathrm{Im}\left[\frac{-1}{\epsilon\left(\mathbf{Q},\omega\right)}\right]
\end{eqnarray}
for the case of HEG. $V_{\chi e}\left(\mathbf{Q}\right)$ is connected
to the relativistic scattering amplitude $\mathcal{M}_{\mathrm{R}}\left(\mathbf{Q}\right)$
in the low-energy limit through the relation 
\begin{eqnarray}
V_{\chi e}\left(\mathbf{Q}\right) & = & \frac{\mathcal{M}_{\mathrm{R}}\left(\mathbf{Q}\right)}{4\,m_{\chi}m_{e}},
\end{eqnarray}
with $m_{\chi}$ being the DM mass. 

It should be noted that, above formalism in terms of the dielectric
function~(i.e., density-density correlation function, see Eq.~(\ref{eq:density-density}))
applies only for a limited set of Lorentz invariant DM interactions~\citep{Trickle:2020oki,Mitridate:2021ctr}.
Because other momentum- or/and angular momentum-dependent effective
operator, as a byproduct of matching onto a non-relativistic effective
field theory from the initial Lorentz invariant interaction, will
also emerge and couple to even more complicated forms of correlation
functions~(e.g., current-current correlation) beyond the density-density
correlation. To be concrete, we consider two simple examples~\citep{Knapen:2021run,Hochberg:2021pkt}
that apply for above discussion: $\mathcal{L}\supset g_{\chi}\phi\bar{\chi}\chi+g_{e}\phi\bar{e}e$,
where a Dirac DM fermion couples to an electron through a scale $\phi$,
and in an analogous fashion, through a vector $V_{\mu}$, i.e., with
Lagrangian $\mathcal{L}\supset g_{\chi}V_{\mu}\bar{\chi}\gamma^{\mu}\chi+g_{e}V_{\mu}\bar{e}\gamma^{\mu}e$.
In these cases, the DM component only couples to the electric density
at the leading order, as the form shown in Eq.~(\ref{eq:density-density}).

Therefore, the DM excitation cross section parallel to Eq.~(\ref{eq:cross_sectioon_crystal})
can be expressed as 
\begin{eqnarray}
\sigma & = & \sum_{\mathbf{G}}\sum_{\mathbf{\mathbf{q}\in1\mathrm{BZ}}}\int\frac{2\,\left|V_{\chi e}\left(\mathbf{\mathbf{G}+\mathbf{q}}\right)\right|^{2}}{v\,V_{\mathrm{Cou}}\left(\mathbf{G}+\mathbf{q}\right)}\,\mathrm{Im}\left[-\epsilon_{\mathbf{G},\mathbf{G}}^{-1}\left(\mathbf{q},\omega\right)\right]\nonumber \\
 &  & \times\delta\left[\frac{\left|\mathbf{G}+\mathbf{q}\right|^{2}}{2m_{\chi}}-\mathbf{v}\cdot\mathbf{\left(\mathbf{G}+\mathbf{q}\right)}+\omega\right]\mathrm{d}\omega.
\end{eqnarray}
For the simplest contact interaction, $V_{\chi e}\left(\mathbf{Q}\right)$
can be replaced by the DM-electron cross section $\sigma_{\chi e}$
as the following, 
\begin{eqnarray}
\left|V_{\chi e}\left(\mathbf{Q}\right)\right|^{2} & = & \frac{\pi\,\sigma_{\chi e}}{\mu_{\chi e}^{2}},\label{eq:contact}
\end{eqnarray}
with $\mu_{\chi e}=m_{e}\,m_{\chi}/\left(m_{e}+m_{\chi}\right)$ being
the reduced mass of the DM-electron pair. Consequently, one obtains
the excitation rate of the electrons in crystalline solid induced
by DM particle as the following: 
\begin{eqnarray}
R & = & \frac{\rho_{\chi}}{m_{\chi}}\left\langle \sigma v\right\rangle \nonumber \\
 & = & \frac{\rho_{\chi}}{m_{\chi}}\frac{\sigma_{\chi e}}{4\,\alpha\,\mu_{\chi e}^{2}}\int\mathrm{d}\omega\int\mathrm{d^{3}}v\,\frac{f_{\chi}\left(\mathbf{v}\right)}{v}\sum_{\mathbf{G}}\sum_{\mathbf{\mathbf{q}\in1\mathrm{BZ}}}\,\left|\mathbf{G}+\mathbf{q}\right|\nonumber \\
 &  & \times\mathrm{Im}\left[-\epsilon_{\mathbf{G},\mathbf{G}}^{-1}\left(\mathbf{q},\omega\right)\right]\Theta\left[v-v_{\mathrm{min}}\left(\left|\mathbf{G}+\mathbf{q}\right|,\,\omega\right)\right],\nonumber \\
\label{eq:eventRate}
\end{eqnarray}
where the bracket $\left\langle \cdots\right\rangle $ denotes the
average over the DM velocity distribution, $\rho_{\chi}=0.3\,\mathrm{GeV/cm^{3}}$
represents the DM local density, and $\Theta$ is the Heaviside step
function, with 
\begin{eqnarray}
v_{\mathrm{min}}\left(\left|\mathbf{G}+\mathbf{q}\right|,\,\omega\right) & = & \frac{\left|\mathbf{G}+\mathbf{q}\right|}{2\,m_{\chi}}+\frac{\omega}{\left|\mathbf{G}+\mathbf{q}\right|}.
\end{eqnarray}
The DM velocity distribution is approximated as a truncated Maxwellian
form in the Galactic rest frame, i.e., $f_{\chi}\left(\mathbf{v}\right)\propto\exp\left[-\left|\mathbf{v}+\mathbf{v}_{\mathrm{e}}\right|^{2}/v_{0}^{2}\right]$$\Theta\left(v_{\mathrm{esc}}-\left|\mathbf{v}+\mathbf{v}_{\mathrm{e}}\right|\right)$,
with the Earth's~(or approximately, the Sun's) velocity $v_{\mathrm{e}}=230\,\mathrm{km/s}$,
the dispersion velocity $v_{0}=220\,\mathrm{km/s}$ and the Galactic
escape velocity $v_{\mathrm{esc}}=544\,\mathrm{km/s}$. If one ignores
the orientation of the crystal structure with respect to the Galaxy
and integrate out the angular parts of the velocity $\mathbf{v}$
and momentum transfer $\left|\mathbf{G}+\mathbf{q}\right|$, the velocity
distribution is considered as isotropic.

Besides, if one assumes the electromagnetic interaction is weak enough
so as to take only the terms up to first order in the resolvent $\left(I-M\right)^{-1}=I+M+M^{2}+\cdots$,
where the identity matrix $I$ and $M$ represent the first and the
the second terms in Eq.~(\ref{eq:RPA matrix}), respectively, the
inverse matrix in Eq.~(\ref{eq:eventRate}) can be approximated as~(see
Eq.~(\ref{eq:RPA matrix})) 
\begin{widetext}
\begin{align}
 & \mathrm{Im}\left[-\epsilon_{\mathbf{G},\mathbf{G}}^{-1}\left(\mathbf{q},\omega\right)\right]\simeq\mathrm{Im}\left[\epsilon_{\mathbf{G},\mathbf{G}}\left(\mathbf{q},\omega\right)\right]\nonumber \\
 & =2\times\frac{4\pi^{2}\alpha}{V\,\left|\mathbf{q}+\mathbf{G}\right|^{2}}\,\sum_{i'}^{c}\sum_{i}^{v}\,\sum_{\mathbf{k}',\mathbf{\mathbf{k}\in1\mathrm{BZ}}}\left|\braket{i'\mathbf{k}'|e^{i\left(\mathbf{\mathbf{\mathbf{q}+\mathbf{G}}}\right)\cdot\hat{\mathbf{x}}}|i\mathbf{k}}\right|^{2}\,\delta\left(\varepsilon_{i'\mathbf{k}'}-\varepsilon_{i\mathbf{k}}-\omega\right),\label{eq:noScreening}
\end{align}
where the Bloch electronic states $\left\{ \ket{i\mathbf{k}}\right\} $
are explicitly labeled with discrete band indices $\left\{ i\right\} $
and crystal momenta $\left\{ \mathbf{k}\right\} $ confined to the
1BZ. Thus above event rate Eq.~(\ref{eq:eventRate}) is explicitly
written as 
\begin{eqnarray}
R & = & \frac{\rho_{\chi}}{m_{\chi}}\frac{\sigma_{\chi e}}{4\,\alpha\,\mu_{\chi e}^{2}}\int\mathrm{d}\omega\int\mathrm{d^{3}}v\,\frac{f_{\chi}\left(\mathbf{v}\right)}{v}\sum_{\mathbf{G}}\sum_{\mathbf{\mathbf{q}\in1\mathrm{BZ}}}\,\left|\mathbf{G}+\mathbf{q}\right|\,\mathrm{Im}\left[-\epsilon_{\mathbf{G},\mathbf{G}}^{-1}\left(\mathbf{q},\omega\right)\right]\,\Theta\left[v-v_{\mathrm{min}}\left(\left|\mathbf{G}+\mathbf{q}\right|,\,\omega\right)\right]\nonumber \\
 & \simeq & \frac{\rho_{\chi}}{m_{\chi}}\frac{2\pi^{2}\sigma_{\chi e}}{\mu_{\chi e}^{2}}\,V\int\mathrm{d^{3}}v\,\frac{f_{\chi}\left(\mathbf{v}\right)}{v}\sum_{\mathbf{G}}\sum_{i'}^{c}\sum_{i}^{v}\int_{1\mathrm{BZ}}\frac{\mathrm{d}^{3}k'}{\left(2\pi\right)^{3}}\int_{1\mathrm{BZ}}\frac{\mathrm{d}^{3}k}{\left(2\pi\right)^{3}}\frac{\left|\int_{\Omega}\mathrm{d}^{3}x\,u_{i'\mathbf{k}'}^{*}\left(\mathbf{x}\right)\,e^{i\mathbf{\mathbf{G}}\cdot\mathbf{x}}\,u_{i\mathbf{k}}\left(\mathbf{x}\right)\right|^{2}}{\left|\mathbf{k}'-\mathbf{k}+\mathbf{G}\right|}\nonumber \\
 &  & \times\Theta\left[v-v_{\mathrm{min}}\left(\left|\mathbf{k}'-\mathbf{k}+\mathbf{G}\right|,\,\varepsilon_{i'\,\mathbf{k}'}-\varepsilon_{i\,\mathbf{k}}\right)\right],\label{eq:R/screening}
\end{eqnarray}
where the periodic wave functions $\left\{ u_{i\mathbf{k}}\left(\mathbf{x}\right)\right\} $
are normalized within the unit cell, over which the integral $\int_{\Omega}\mathrm{d}^{3}x\,\left(\cdots\right)$
is performed. It is straightforward to verify that the event rate
in Eq.~(\ref{eq:R/screening}) exactly corresponds to the case without
the screening effect~\citep{Essig:2015cda}.
\end{widetext}

\subsection{\label{sub:screening_semiconductor}Screening effect in DM direct
detection}

Now we put above formulas into practical computations. We will concretely
calculate the screening effect on sensitivities of diamond- and silicon-base
detectors to the Galactic DM halo, discussing the local field effects
in different computational approaches, and compare our results with
those calculated with the $\mathtt{GPAW}$ code~\citep{Knapen:2021run}.

\begin{figure*}[t]
\begin{centering}
\includegraphics[scale=0.64]{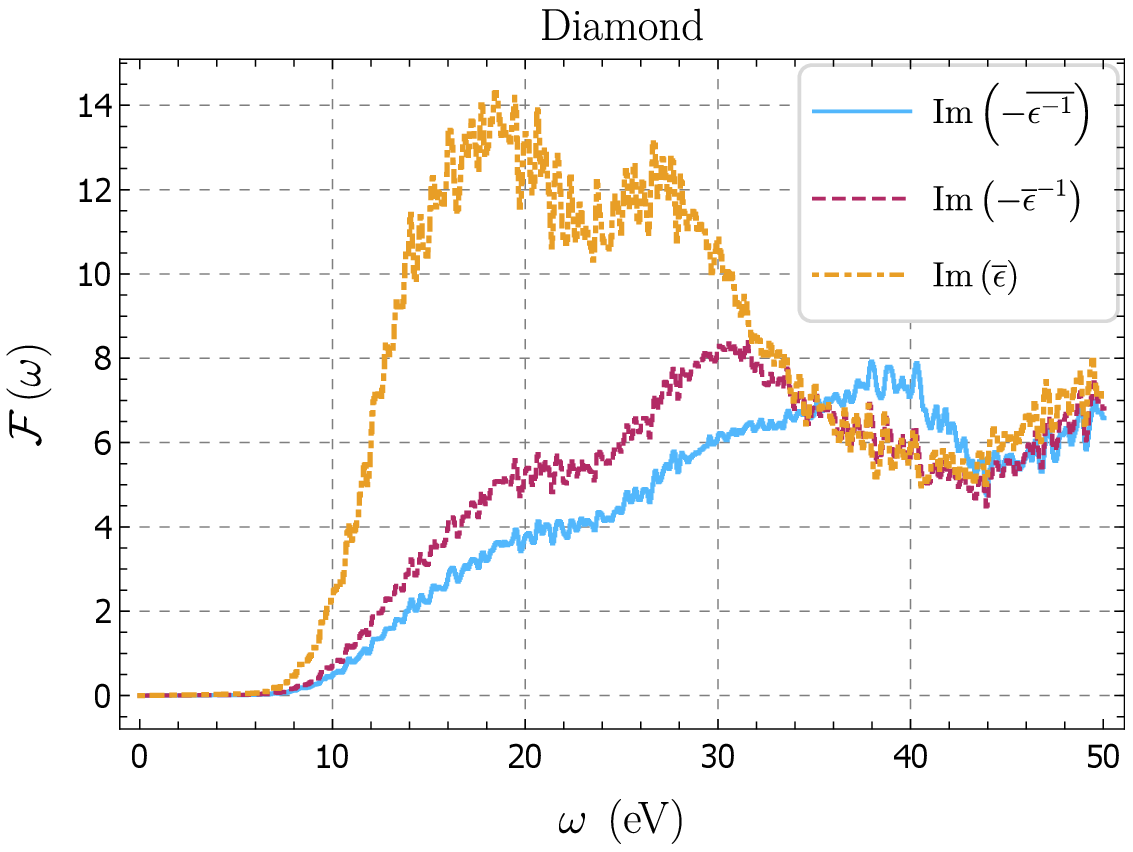}\hspace{0.8cm}\includegraphics[scale=0.64]{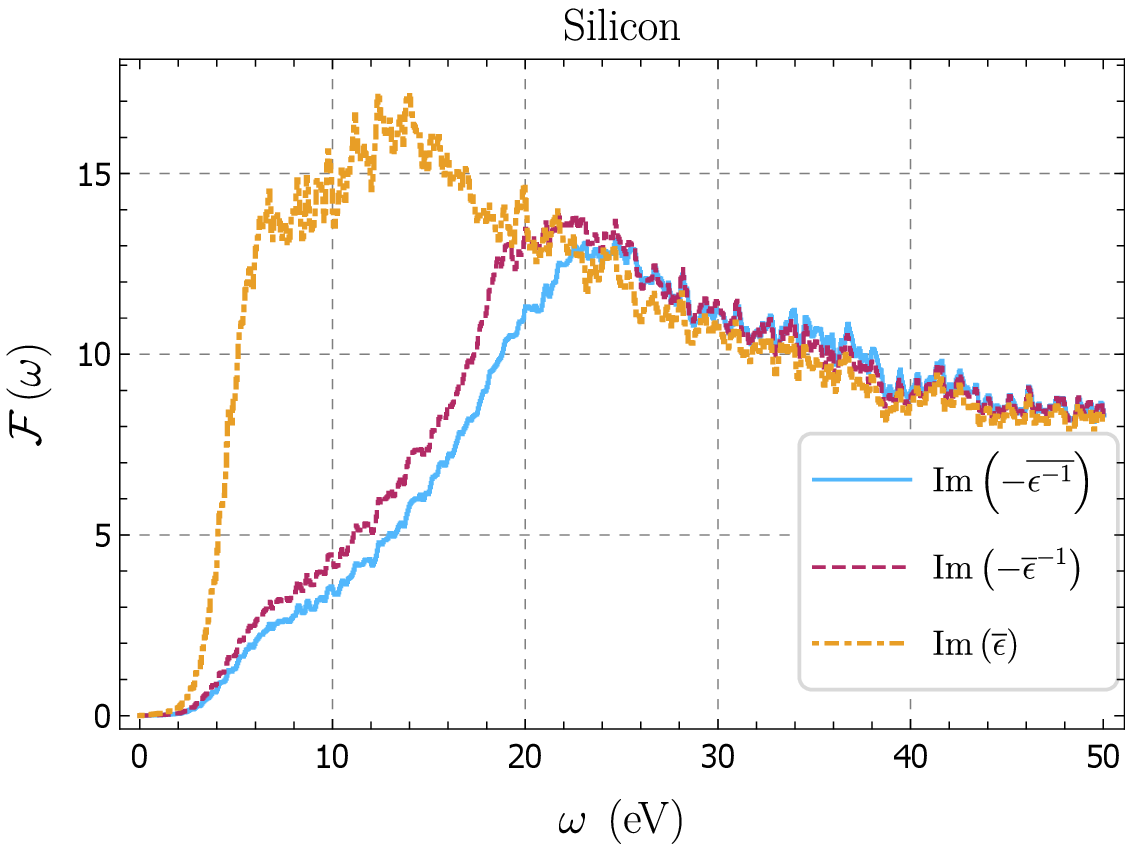}\vspace{0.5cm}
\par\end{centering}
\begin{centering}
\includegraphics[scale=0.64]{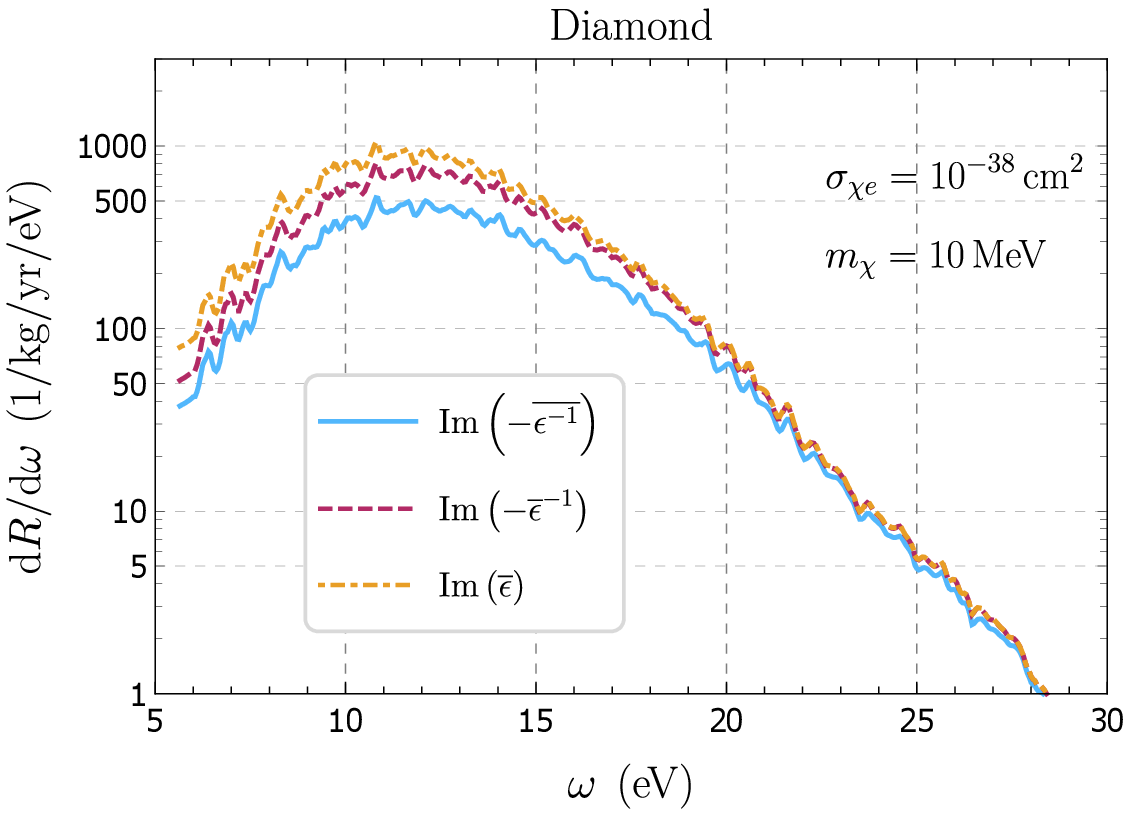}\hspace{0.8cm}\includegraphics[scale=0.64]{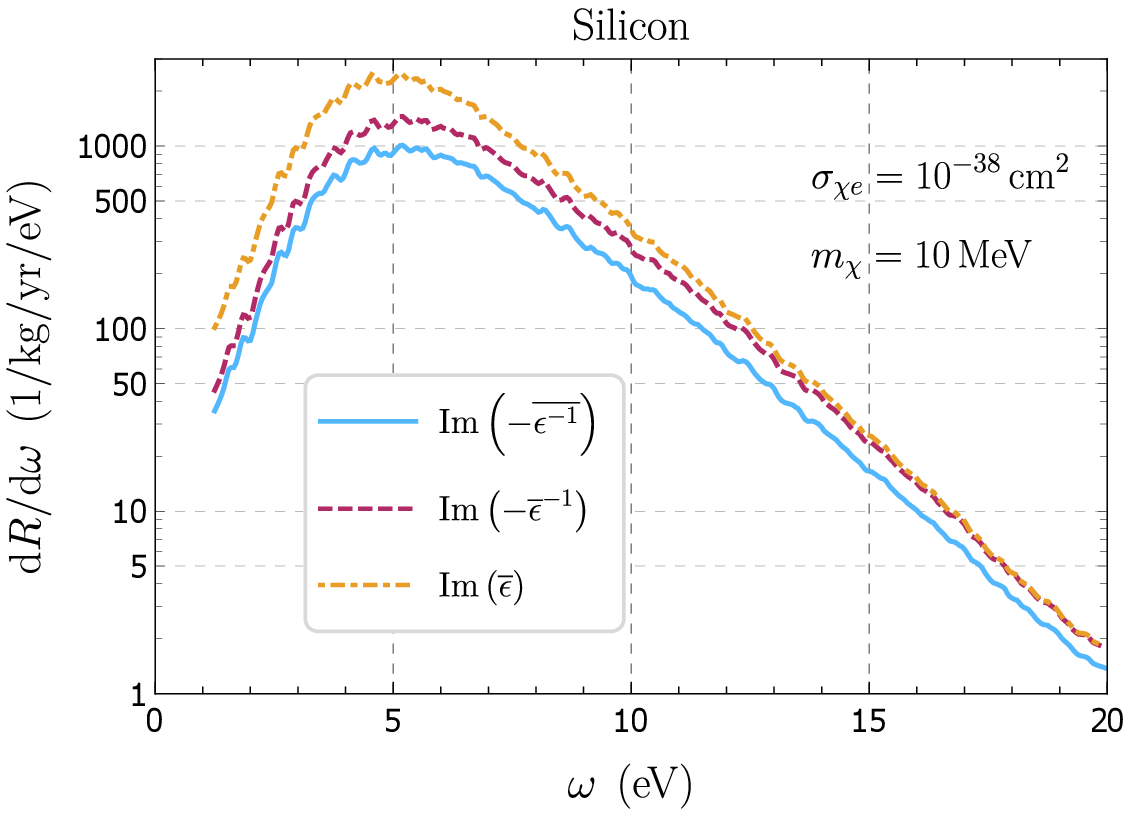}
\par\end{centering}
\caption{\label{fig:comparison}\textbf{\textit{Top}}: The nondimensional factor
$\mathcal{F}\left(\omega\right)$ defined in Eq.~(\ref{eq:factor})
for the inverse dielectric functions $\mathrm{Im}\left[-\overline{\epsilon^{-1}}\right]$~(\textit{blue
solid}), $\mathrm{Im}\left[-\overline{\epsilon}^{-1}\right]$~(\textit{red
dashed}) and $\mathrm{Im}\left[\overline{\epsilon}\right]$~(\textit{orange
dot-dashed}) for diamond~(\textit{left}) and silicon~(\textit{right}),
respectively. \textbf{\textit{Bottom}}: The screening effect on differential
rate spectra in diamond~(\textit{left}) and silicon targets~(\textit{right}),
for a $10\,\mathrm{MeV}$ DM particle and cross section $\sigma_{\chi e}=10^{-38}\,\mathrm{cm}^{2}$,
respectively. See text for details.}
\end{figure*}

In practical computation, it is convenient to reinterpret the integration
over the momenta $\mathbf{G}$ and $\mathbf{q}$ in Eq.~(\ref{eq:eventRate})
in terms of variable $\left|\mathbf{q}+\mathbf{G}\right|$. To this
end, we first calculate the angular-averaged inverse dielectric function~\citep{Knapen:2021run}:
\begin{eqnarray}
\overline{\epsilon^{-1}}\left(Q,\omega\right) & \equiv & \frac{1}{N\left(Q\right)}\sum_{\mathbf{q},\mathbf{G}}\epsilon_{\mathbf{G},\mathbf{G}}^{-1}\left(\mathbf{q},\omega\right)\delta_{Q,\,\left|\mathbf{q}+\mathbf{G}\right|},\label{eq:inverse_definition}
\end{eqnarray}
where $N\left(Q\right)\equiv\sum_{\mathbf{q},\mathbf{G}}\delta_{Q,\,\left|\mathbf{q}+\mathbf{G}\right|}$,
and $Q$ is an arbitrary transferred momentum beyond the 1BZ. Note
that this definition takes into account the LFEs. As a consequence,
the excitation rate in Eq.~(\ref{eq:eventRate}) can be equivalently
recast as 
\begin{align}
 & R=\frac{\rho_{\chi}}{m_{\chi}}\frac{\sigma_{\chi e}\,N_{\mathrm{cell}}}{4\,\alpha\,\mu_{\chi e}^{2}}\int\mathrm{d}\omega\int\mathrm{d^{3}}v\,\frac{f_{\chi}\left(\mathbf{v}\right)}{v}\int\frac{\Omega\,\mathrm{d}^{3}Q}{\left(2\pi\right)^{3}}\nonumber \\
 & \times Q\,\mathrm{Im}\left[-\overline{\epsilon^{-1}}\left(Q,\omega\right)\right]\,\Theta\left[v-v_{\mathrm{min}}\left(Q,\,\omega\right)\right],\label{eq:EventRateELF}
\end{align}
where $N_{\mathrm{cell}}$ is the number of the unit cells in the
target material. In addition, there is an alternative definition of
the ELF~\citep{Knapen:2021run}, where the inverse dielectric function
is obtained by first calculating the directionally averaged dielectric
function 
\begin{eqnarray}
\overline{\epsilon}\left(Q,\omega\right) & \equiv & \frac{1}{N\left(Q\right)}\sum_{\mathbf{q},\mathbf{G}}\epsilon_{\mathbf{G},\mathbf{G}}\left(\mathbf{q},\omega\right)\delta_{Q,\,\left|\mathbf{q}+\mathbf{G}\right|},\label{eq:averaged dielectric}
\end{eqnarray}
and then approximating the the inverse dielectric function as $\mathrm{Im}\left[-\overline{\epsilon^{-1}}\left(Q,\omega\right)\right]\simeq\mathrm{Im}\left[-\overline{\epsilon}^{-1}\left(Q,\omega\right)\right]$.
This approximation neglects the LFEs because only the information
of the diagonal terms of matrix $\epsilon_{\mathbf{G},\mathbf{G}}\left(\mathbf{q},\omega\right)$
is retained in ELF. In a similar manner, the inverse dielectric function
for the non-screening case in Eq.~(\ref{eq:noScreening}) can be
approximated as $\mathrm{Im}\left[-\overline{\epsilon^{-1}}\left(Q,\omega\right)\right]\simeq\mathrm{Im}\left[\overline{\epsilon}\left(Q,\omega\right)\right]$.
So one of our purpose is to investigate the in-medium screening effect
of the DM-electron excitation process. Besides, it is interesting
to compare the results drawn form the two definitions of the inverse
dielectric functions. While the definition Eq.~(\ref{eq:inverse_definition})
faithfully reproduces the definition in Eq.~(\ref{eq:eventRate}),
the LFEs are neglected in Eq.~(\ref{eq:averaged dielectric}). To
explore the consequence of the LFEs, we concretely compute the excitation
rates for diamond and silicon targets based on Eq.~(\ref{eq:inverse_definition}),
and make a comparison with the results obtained from the definition
Eq.~(\ref{eq:averaged dielectric}).

\begin{figure*}
\begin{centering}
\includegraphics[scale=0.7]{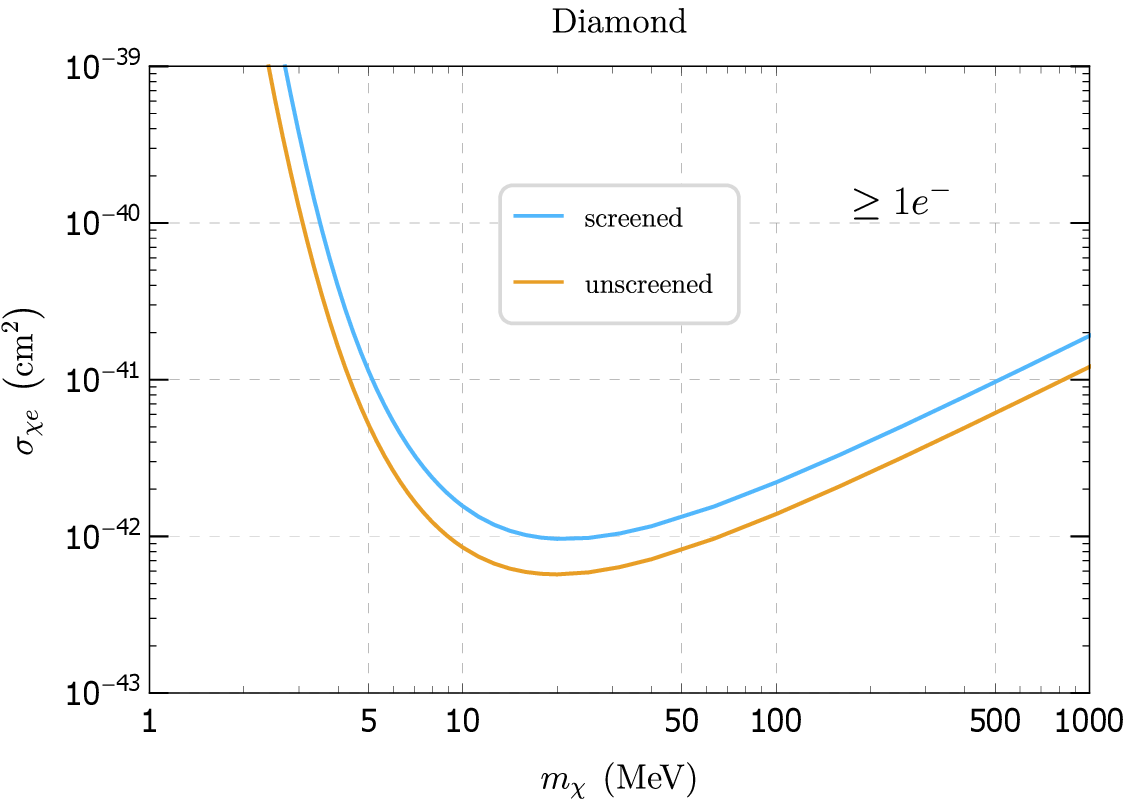}\hspace{0.3cm}\includegraphics[scale=0.7]{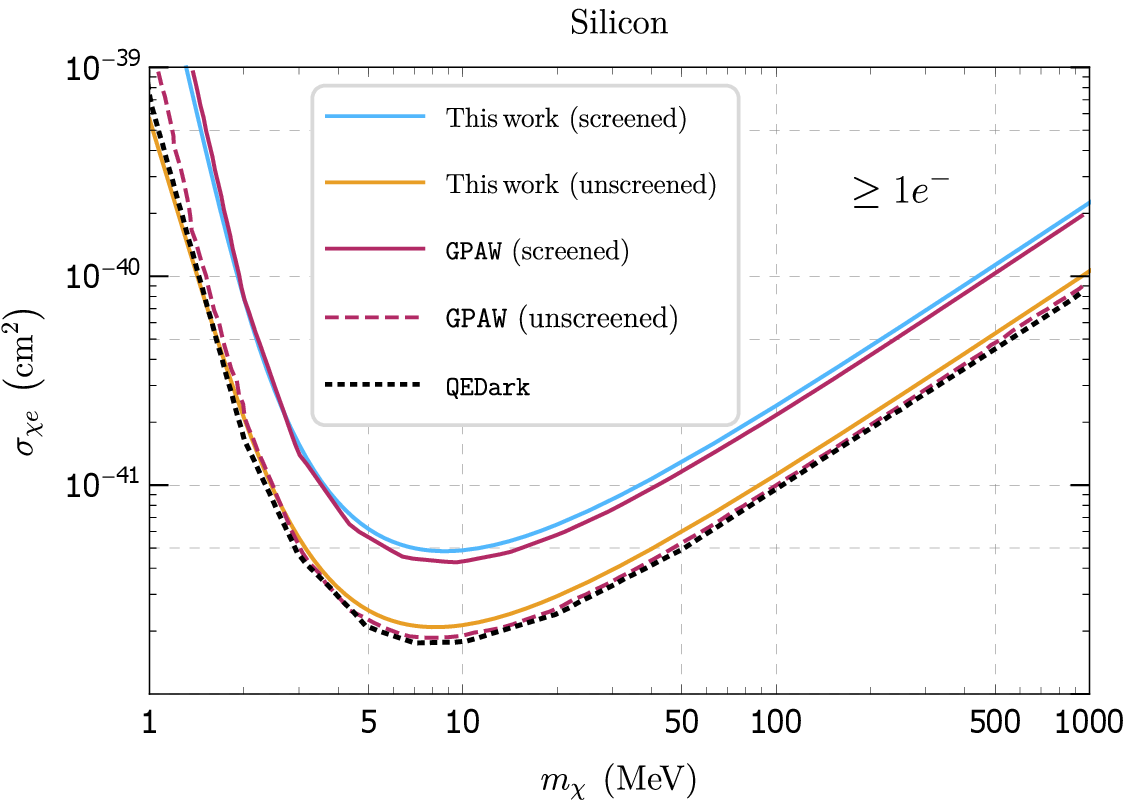}\vspace{0.3cm}
\par\end{centering}
\begin{centering}
\includegraphics[scale=0.7]{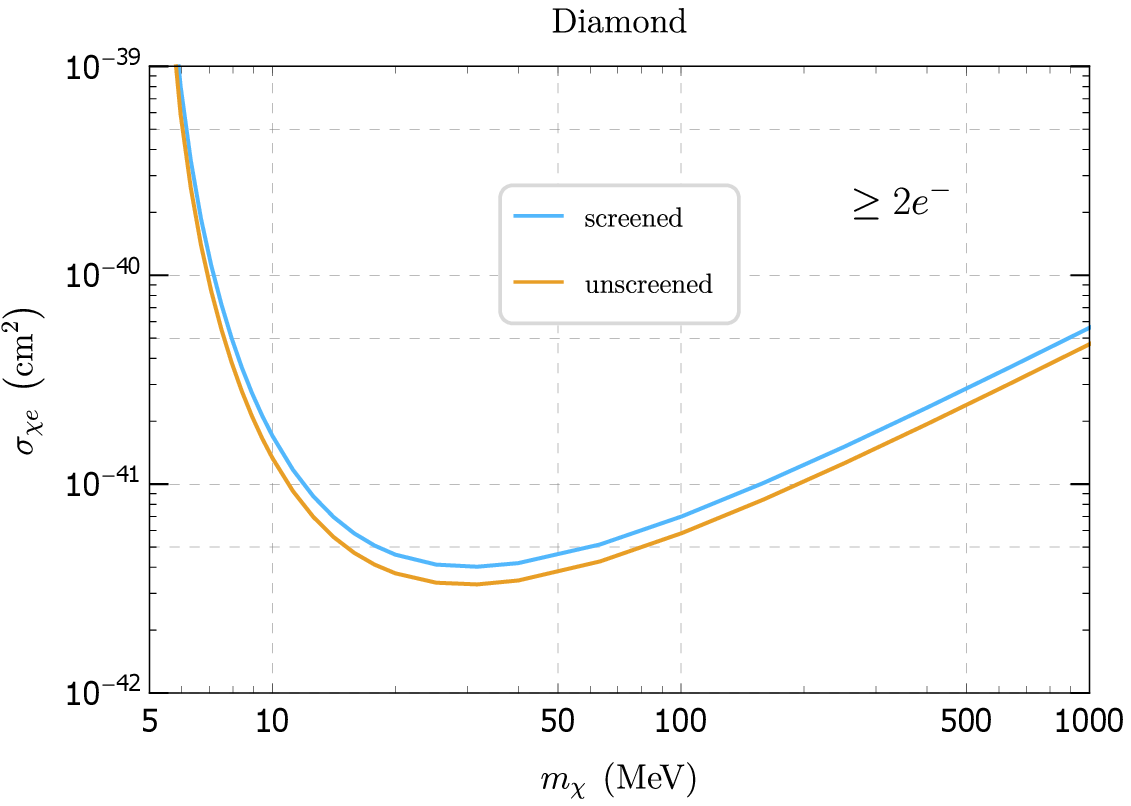}\hspace{0.3cm}\includegraphics[scale=0.7]{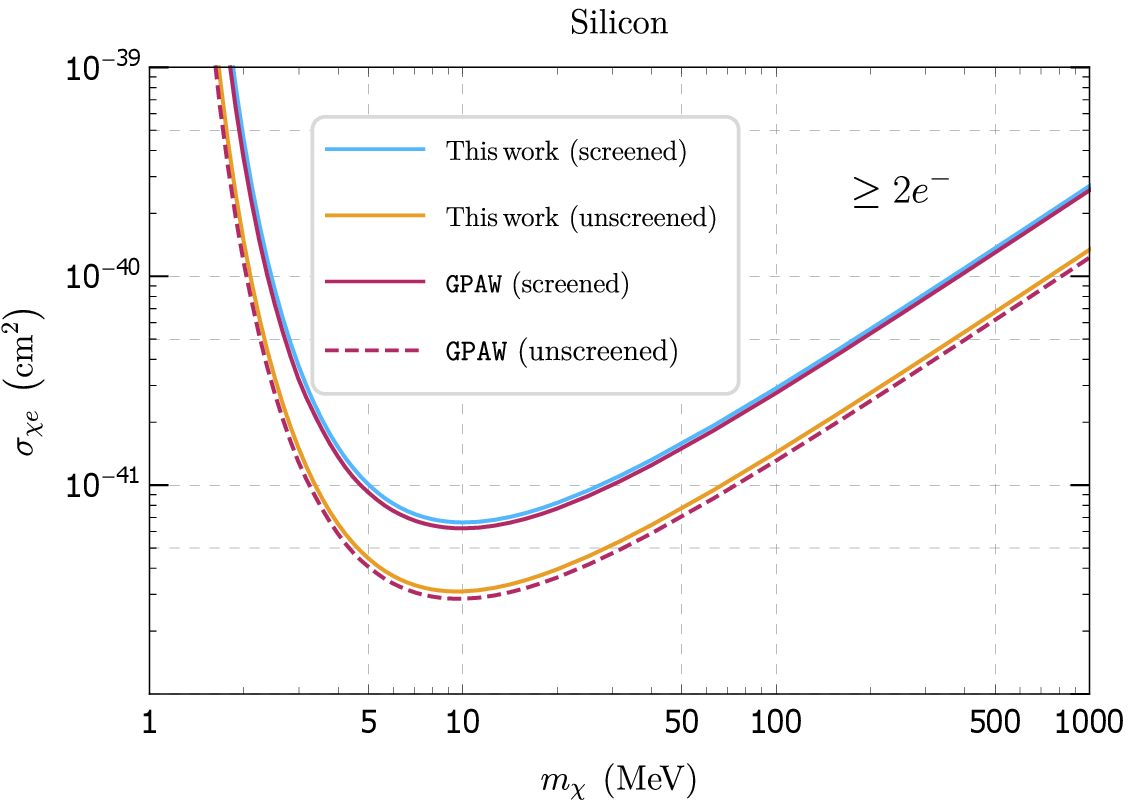}
\par\end{centering}
\caption{\textit{\label{fig:Sensitivities}}95\% C.L. constraints on the DM-electron
cross section $\sigma_{\chi e}$ with~(\textit{blue}) and without~(\textit{orange})
the screening effect for the diamond target, assuming $1e^{-}$~(\textbf{\textit{upper
left}}) and $2e^{-}$(\textbf{\textit{bottom left}}) thresholds.\textit{
}\textbf{\textit{Upper right}}: Sensitivities of various calculations~for
silicon target at the 95\% C.L.: with~(\textit{blue}) and without~(\textit{orange})
screening effect in this work, $\mathtt{GPAW}$ estimation~\citep{Knapen:2021run}
for screened~(\textit{red solid}) and unscreened cases~(\textit{red
dashed}), and $\mathtt{QEDark}$~\citep{Essig:2015cda}~(\textit{black
dotted}), respectively. All constraints are calculated with negligible
background and a 1kg-year exposure. For comparison purpose, the DM
distribution parameters $\rho_{\chi}=0.4\,\mathrm{GeV/cm^{3}}$, $v_{\mathrm{e}}=240\,\mathrm{km/s}$,
$v_{0}=230\,\mathrm{km/s}$ and $v_{\mathrm{esc}}=600\,\,\mathrm{km/s}$
are adopted in alignment. \textbf{\textit{Bottom right}}: 95\% C.L
$2e^{-}$ threshold sensitivities for silicon target: with~(\textit{blue})
and without~(\textit{orange}) screening effect in this work, $\mathtt{GPAW}$
estimations~\citep{Knapen:2021run} for screened~(\textit{red solid})
and unscreened cases~(\textit{red dashed}), respectively. See text
for details.}
\end{figure*}

Using $\mathtt{Quantum\:Espresso}$ package~\citep{Giannozzi_2009}
plus a norm-conserving pseudopotential~\citep{PhysRevLett.43.1494},
we perform the DFT calculation to obtain the Bloch eigenfunctions
and eigenvalues using the local-density approximation~\citep{PhysRevB.23.5048}
for the exchange-correlation functional, on a uniform 6$\times$6$\times$6~(5$\times$5$\times$5)
$k$-point mesh for diamond (silicon) via the Monkhorst-Pack~\citep{PhysRevB.13.5188}
scheme. A core cutoff radius of $1.3\,\mathrm{Bohr}$~($1.8\,\mathrm{Bohr}$)
is adopted and the outermost four electrons are treated as valence
for both diamond and silicon. The energy cut $\varepsilon_{\mathrm{cut}}$
is set to $200\,\mathrm{Ry}$~($70\,\mathrm{Ry}$) and lattice constant
3.577~Å~~(5.429~Å) for diamond (silicon) obtained from experimental
data is adopted. The matrix $\epsilon_{\mathbf{G},\mathbf{G}'}^{-1}$
is calculated via directly inverting the matrix Eq.~(\ref{eq:RPA matrix})
at the RPA level with the $\mathtt{YAMBO}$ package~\citep{Marini2009.yambo},
with a matrix cutoff of $50\,\mathrm{Ry}$~($20\,\mathrm{Ry}$),
corresponding to $Q\leq30\,\mathrm{keV}$~($20\,\mathrm{keV}$) for
diamond (silicon). An energy bin width $\Delta\omega=0.05\,\mathrm{eV}$
is adopted within the range from $0$ to $50\,\mathrm{eV}$.

In order to gauge the screening effect and the difference between
the two ELFs, we introduce the following nondimensional factor and
present it in the upper row of Fig.~\ref{fig:comparison}, 
\begin{eqnarray}
\mathcal{F}\left(\omega\right) & = & \sum_{\mathbf{G}}\int_{1\mathrm{BZ}}\frac{\varOmega\,\mathrm{d}^{3}q}{\left(2\pi\right)^{3}}\,\mathrm{Im}\left[-\epsilon_{\mathbf{G},\mathbf{G}}^{-1}\left(\mathbf{q},\omega\right)\right]\nonumber \\
 & = & \int\frac{\varOmega\,\mathrm{d}^{3}Q}{\left(2\pi\right)^{3}}\,\mathrm{Im}\left[-\overline{\epsilon^{-1}}\left(Q,\omega\right)\right],\label{eq:factor}
\end{eqnarray}
for the case of unscreened, screened with LFEs and screened without
LFEs, respectively. While it is evident from the top panel of Fig.~\ref{fig:comparison}
that the screening effect is remarkable in the low-energy regime~($\omega\lesssim30\,\mathrm{eV}$
for diamond and $\omega\lesssim20\,\mathrm{eV}$ for silicon), the
factor $\mathcal{F}\left(\omega\right)$ calculated from the dielectric
function $\mathrm{Im}\left[-\overline{\epsilon^{-1}}\right]$ in Eq.~(\ref{eq:inverse_definition})
differs from the one computed from $\mathrm{Im}\left[-\overline{\epsilon}^{-1}\right]$
below Eq.~(\ref{eq:averaged dielectric}) by a factor less than 50\%
in relevant energy range. In this sense, the dielectric function $\mathrm{Im}\left[-\overline{\epsilon^{-1}}\left(Q,\omega\right)\right]\simeq\mathrm{Im}\left[-\overline{\epsilon}^{-1}\left(Q,\omega\right)\right]$
amounts to an acceptable approximation. In the energy range ($\omega>30\,\mathrm{eV}$
for diamond and $\omega>20\,\mathrm{eV}$ for silicon), the screening
effect turns negligible. In the bottom panel of Fig.~\ref{fig:comparison}
we present the corresponding differential spectra for diamond~(left)
and silicon~(right) for a DM mass $m_{\chi}=10\,\mathrm{MeV}$ and
a benchmark cross section $\sigma_{\chi e}=10^{-38}\,\mathrm{cm}^{2}$,
respectively.

To translate the spectrum into excited electron signals, we adopt
the model~\citep{Essig:2015cda} where the secondary electron-hole
pairs triggered by the primary one are described with the mean energy
per electron-hole pair $\varepsilon$ in high energy recoils. In this
picture, the ionization charge $\mathcal{Q}$ is then given by 
\begin{eqnarray}
\mathcal{Q}\left(\omega\right) & = & 1+\left\lfloor \left(\omega-E_{g}\right)/\varepsilon\right\rfloor ,
\end{eqnarray}
where $\left\lfloor x\right\rfloor $ rounds $x$ down to the nearest
integer, and $E_{g}$ denotes the band gap. Thus, from the energy
spectra we estimate the sensitivities of a 1 kg-yr diamond~(silicon)
detector in Fig.~\ref{fig:Sensitivities}, adopting a band gap value
$E_{g}=5.47\,\mathrm{eV}$~($1.12\,\mathrm{eV}$) and assuming an
average energy $\varepsilon=13\,\mathrm{eV}$~($3.6\,\mathrm{eV}$)
for producing one electron-hole pair for diamond~\citep{Kurinsky:2019pgb}~(silicon~\citep{Essig:2015cda}).
In the left panel shown are the 95\% C.L. constraints for diamond
target with a kg-year exposure for the screened and unscreened cases,
assuming $1e^{-}$~(top left) and $2e^{-}$~(bottom left) thresholds.
Compared to the $1e^{-}$ threshold, the discrepancy between the screened
and unscreened estimations narrows, which can be attributed to the
large $\varepsilon=13\,\mathrm{eV}$ that pushes relevant energy into
the regime where screening effect begins to wear off. Besides, in
order to make comparison with previous $\mathtt{GPAW}$~\citep{Knapen:2021run}
and $\mathtt{QEDark}$~\citep{Essig:2015cda} calculations, we present
in the right panel in Fig.~\ref{fig:Sensitivities} the 95\% C.L.
kg-year exposure projected sensitivities for silicon target with a
single electron threshold and no background events. In practical evaluation
of dielectric matrix Eq.~(\ref{eq:RPA matrix}), a small broadening
parameter $\eta=0.1\,\mathrm{eV}$ is adopted for both diamond and
silicon, instead of an infinitesimal energy width $0^{+}$. A non-vanishing
$\eta$ usually brings a long tail extending into the gap region,
and hence induces a small contribution to the excitation rate around
$E_{g}$. Theoretically, the smaller the parameter $\eta$, the more
accurate the computation is, but on the other hand, a smaller $\eta$
also requires a finer energy width $\Delta\omega$ and a denser $k$-point
mesh to smear the spectra. As pointed out in Ref.~\citep{Knapen:2021bwg},
there are expected to be $\mathcal{O}\left(1\right)$ uncertainties
in the energy range $\omega\lesssim2E_{g}$, due to the strong fluctuations.
While the event rates calculated in this work generally coincide well
with the $\mathtt{GPAW}$ results, the latter give a more conservation
estimation in the low-energy region for $1e^{-}$ threshold, as a
result of different choices of parameter $\eta$. Such uncertainties
do not cause a severe problem because they mainly occur in the region
plagued by a large noises in most detectors~(in the single-electron
bin, for instance), and thus are usually excluded from most experimental
analyses. If a $2e^{-}$ threshold is adopted, the $\mathtt{YAMBO}$
and $\mathtt{GPAW}$ calculations well coincide in the whole DM mass
range, which is clearly seen from the bottom right panel of Fig.~\ref{fig:Sensitivities}.

\section{\label{sec:Solar_Reflection}Light DM particles reflected from the
Sun}

The idea of detecting the MeV-scale DM particles via solar reflection
was first proposed in Ref.~\citep{An:2017ojc}, and is further discussed
in Refs.~\citep{Chen:2020gcl,Emken:2021lgc}\footnote{{\selectfont Ref.~\citep{Emken_2018} developed this idea simultaneously
in the context of nuclear interactions, and similar proposal of detecting
solar DM particles from the evaporation effect can be traced back
to an earlier work~\citep{Kouvaris:2015nsa}.}}. In this study, our interest is focused on the case where the DM
component couples to electron via a heavy mediator, for both scalar~($S\otimes S$)
and vector~($V\otimes V$) interaction types\footnote{{\selectfont These simple models are subject to severe cosmological
constraints from BBN and CMB below a few MeV. See Refs.~\citep{lin2019tasi,kahn2021searches}
for a review. }}, which corresponds to the contact interaction Eq.~(\ref{eq:contact}). 

Previous investigation~\citep{Kopp:2009et} pointed out that even
for these leptophilic DM models, the effective DM-nucleon cross section
arising from lepton-assisted loop-induced processes may compete or
even overwhelm that of the DM-electron interaction in DM direct detection
experiments. However, a closer analysis shows that, the DM-nucleus
cross section in the Sun is so significantly suppressed by a much
lighter DM mass~(MeV), and a smaller charge number $Z$~(predominantly
hydrogen and helium in the Sun) compared to the scenario in direct
detection, that it can be safely neglected compared to the DM-electron
scattering\footnote{{\selectfont A detailed discussion is arranged in Appendix~\ref{sec:AppendixA}.}}. 

Here we begin with a short review of related physics in the Sun and
explain in detail the methodology we adopt in this paper. As in Refs.~\citep{An:2017ojc,Emken:2021lgc},
in this work we also take a Monte Carlo simulation approach to describe
the solar reflection of the DM particles. Then we generalize discussion
in previous sections to the case of solar-reflected DM flux, specifying
the screening effect on relevant detection experiments under way,
and in plan for the near future.

\subsection{Initial condition}

The standard description of the DM's encounter with the Sun has been
well established in the literature~\citep{Gould:1987ir,Gould:1987ww,Gould:1991hx},
which provides an elegant analytic approach in dealing with DM capture
and evaporation process. Related arguments can be applied to the present
discussion. The starting point of our discussion is the rate $\Gamma$
at which the DM flux reaches the solar surface, which is given by\footnote{{\selectfont We provide a concise derivation of this expression
in Appendix~\ref{sec:Appendix}.}}:
\begin{eqnarray}
\Gamma\left(m_{\chi}\right) & = & \frac{\rho_{\chi}}{m_{\chi}}\pi\int\frac{f_{\chi}\left(\mathbf{u}\right)}{u}\,\mathrm{d}^{3}u\int_{0}^{w^{2}\left(u,R_{\odot}\right)R_{\odot}^{2}}\mathrm{d}J^{2}\nonumber \\
 & = & \frac{\rho_{\chi}}{m_{\chi}}\pi R_{\odot}^{2}\int\frac{w^{2}\left(u,R_{\odot}\right)}{u}f_{\chi}\left(\mathbf{u}\right)\mathrm{d}^{3}u\int_{0}^{1}\mathrm{d}\left(\sin^{2}\theta\right)\nonumber \\
 & = & \frac{\rho_{\chi}}{m_{\chi}}\pi R_{\odot}^{2}\int\left[\frac{u^{2}+v_{\mathrm{esc}}^{2}\left(R_{\odot}\right)}{u}\right]f_{\chi}\left(\mathbf{u}\right)\mathrm{d}^{3}u,\label{eq:SurfaceRate}
\end{eqnarray}
where $R_{\odot}$ is the radius of the Sun, $J$ represents the angular
momentum of the DM particle in the solar central field, and $w^{2}\left(u,R_{\odot}\right)=u^{2}+v_{\mathrm{esc}}^{2}\left(R_{\odot}\right)$,
with $v_{\mathrm{esc}}\left(r\right)$ being the solar escape velocity
at radial distance $r$.
\begin{figure}
\begin{centering}
\includegraphics[scale=0.35]{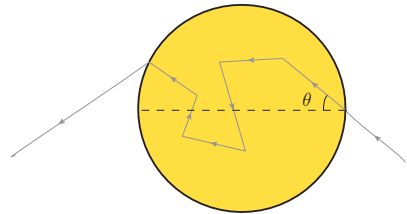}
\par\end{centering}
\caption{\label{fig:Illustration}The DM particle enters the body of the sun
with an angle $\theta$, collides with solar electrons and finally
escapes from the Sun. For the purpose of illustration, the DM trajectory
is projected onto a single plane.}
\end{figure}

Instead of shooting the sampled particles from a large distance with
an impact parameter~\citep{An:2017ojc,Emken:2021lgc}, we inject
them at the surface of the Sun by using the second line in Eq.~(\ref{eq:SurfaceRate})
as the initial condition of the impinging DM flux, which is more convenient
if the DM distribution $f_{\chi}\left(\mathbf{u}\right)$ is approximated
as isotropic. On one hand, the incident velocity at the surface $w$
is connected to the halo velocity with $\sqrt{u^{2}+v_{\mathrm{esc}}^{2}\left(R_{\odot}\right)}$;
on the other, the angle $\theta$ between the incident and the solar
radial directions can be determined by angular momentum $J$, i.e.,
$J^{2}=w^{2}R_{\odot}^{2}\sin^{2}\theta$. So for each particular
injection speed $w\left(u,R_{\odot}\right)$~(with weighting factor
$\sim w^{2}\left(u,R_{\odot}\right)f_{\chi}\left(u\right)\mathrm{d}u/u$)\footnote{{\selectfont Notice that $w$ and $u$ are in one-to-one correspondence.
Here $f_{\chi}\left(u\right)=\int u^{2}f_{\chi}\left(\mathbf{u}\right)\mathrm{d}\varOmega$,
i.e., the DM speed distribution after the directions are integrated
out. }}, the directions of the DM particles at surface are sampled evenly
in $\sin^{2}\theta$. Fig.~\ref{fig:Illustration} shows a schematic
sketch of the initial condition for the simulation.

\subsection{Propagation in the Sun}

Then the trajectories of these sampled DM particles are simulated.
To be specific, once a DM particle enters the bulk of the Sun, we
first determine whether it will collide with surrounding electrons
in the next time step $\Delta t$, which is described with the probability
\begin{eqnarray}
P_{\mathrm{collision}} & = & 1-\exp\left[-\lambda(t)\,\Delta t\right],\label{eq:collisionP}
\end{eqnarray}
where 
\begin{eqnarray}
\lambda\left(t\right) & = & n_{e}\left(r\right)\left\langle \sigma_{\chi e}\cdot\left|\mathbf{w}-\mathbf{u}_{e}\right|\right\rangle \nonumber \\
 & = & n_{e}\left(r\right)\,\sigma_{\chi e}\left[\frac{u_{0}}{\sqrt{\pi}}\,\exp\left(-w^{2}/u_{0}^{2}\right)\right.\nonumber \\
 &  & \left.+\left(w+\frac{u_{0}^{2}}{2\,w}\right)\mathrm{erf}\left(\frac{w}{u_{0}}\right)\right]\label{eq:lambda}
\end{eqnarray}
is implicitly dependent on temporal parameter $t$, $\left\langle \cdots\right\rangle $
denotes the average over the relative velocity $\mathbf{w}-\mathbf{u}_{e}$
between the DM particle and the surrounding electrons, and $n_{e}\left(r\right)$
is local electron number density. The Maxwellian distribution $f_{e}\left(\mathbf{u}_{e}\right)$
is explicitly written as 
\begin{eqnarray}
f_{e}\left(\mathbf{u}_{e}\right) & = & \left(\sqrt{\pi}u_{0}\right)^{-3}\exp\left(-\frac{u_{e}^{2}}{u_{0}^{2}}\right),
\end{eqnarray}
where $u_{0}=$ $\sqrt{2\,T_{\odot}\left(r\right)/m_{e}}$, and \textcolor{black}{$T_{\odot}\left(r\right)$
}is\textcolor{black}{{} the local temperature. }

Next, a random number $\xi$ between $0$ and $1$ is generated. If
$\xi>P_{\mathrm{collision}}$ we conclude that a scattering event
will not happen, and the DM particle propagates to the next location.
The gravitational field can be specified by referring to the Standard
Sun Model (SSM)~AGSS09~\citep{Serenelli:2009yc}. The number density
of the ionized electrons is determined by the condition of charge
neutrality~\citep{Emken:2021lgc}. If $\xi<P_{\mathrm{collision}}$,
on the other hand, a scattering event is assumed to occur. In this
case, further random numbers are generated to pick out the velocity
of the electron participating in the collision, as well as the scattering
angles in the center-of-mass frame, so that the outgoing state of
the scattered DM particle can be determined after a coordinate transformation
back to the solar reference~\citep{Liang:2018cjn}.

Then this simulation process continues until one of the following
two conditions is satisfied: (1) DM particle reaches the solar surface
with a velocity $w>v_{\mathrm{esc}}\left(R_{\odot}\right)$~(if not
so, the direction of the velocity is flipped, and the simulation continues
until next time the DM particle reaches the surface, and this process
is repeated again.); (2) the DM particle is regarded as captured.
While the first criterion is straightforward in practice, the second
is not so definite, especially considering that a temporarily trapped
sub-MeV DM particle is so volatile that after a few collisions it
will be kicked out of the solar gravitational well, namely, be evaporated.
In this case, the boundary between evaporation and reflection no longer
exists, and one should describe them in a unified approach. As will
be explained in the following discussions, in practice we specify
the criterion for capture such that the DM particle scatters more
than 200 times in optically thick regime~(i.e., $\sigma_{\chi e}\geq10^{-36}\,\mathrm{cm}^{2}$
in our practical computation).
\begin{figure*}
\begin{centering}
\includegraphics[scale=0.52]{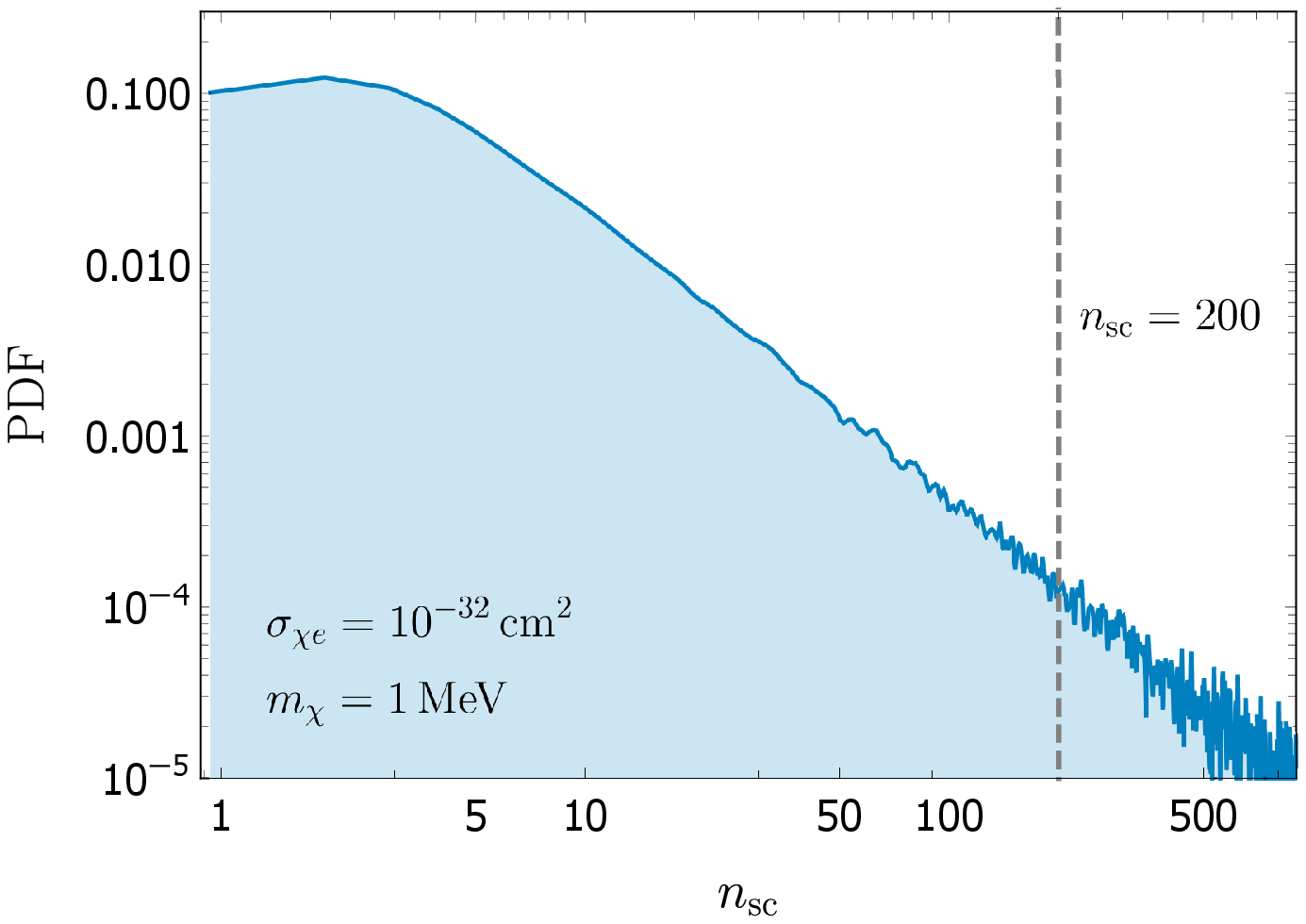}\hspace{1cm}\includegraphics[scale=0.684]{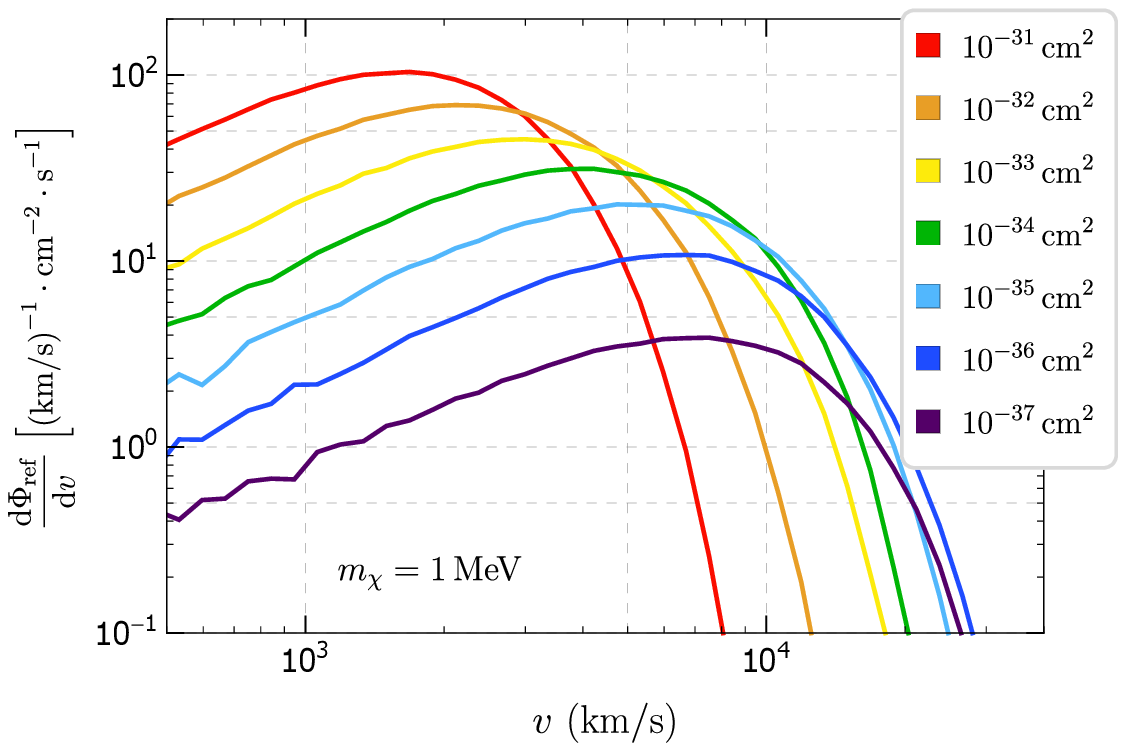}
\par\end{centering}
\caption{\label{fig:VelocitySpectrum}\textbf{\textit{Left}}: The PDF of scattering
number $n_{\mathrm{sc}}$ in simulations, for a DM particle with mass
$m_{\chi}=1\,\mathrm{MeV}$ and DM-electron cross section $\sigma_{\chi e}=1\times10^{-32}\,\mathrm{cm}^{2}.$
Events undergoing more than $200$ collisions are categorized as capture
events. \textbf{\textit{Right}}: The reflected DM differential flux
$\mathrm{d}\Phi_{\mathrm{ref}}/\mathrm{d}v$ of a $1\,\mathrm{MeV}$
DM particle with various representative cross sections ranging from
$1\times10^{-37}\,\mathrm{cm}^{2}$ to $1\times10^{-31}\,\mathrm{cm}^{2}$,
respectively. See text for details.}
\end{figure*}

\subsection{\label{sub:Velocity-distribution}Spectrum of reflection flux}

As DM particle reaches the solar surface, we find out whether it has
ever suffered a collision. If not, the sample is categorized as the
galactic background and thus is taken out of the tally. If the ongoing
DM particle has been scattered more than once and leaves with a velocity
$w$ greater than the escape velocity at surface $v_{\mathrm{esc}}\left(R_{\odot}\right)$,
this velocity is red-shifted such that $v=\sqrt{w^{2}-v_{\mathrm{esc}}^{2}\left(R_{\odot}\right)+v_{\mathrm{esc}}^{2}\left(D\right)}$
(with $D$ being the Earth-Sun distance), and is put into the prepared
bins for the velocity spectrum at the terrestrial detectors. For those
leave the Sun with a velocity $w<v_{\mathrm{esc}}\left(R_{\odot}\right)$,
we consider them as captured.

When a DM particle dives into the Sun, it may be kicked out after
a few collisions, or may be confined to the solar gravitational field
for a long time. In the latter case, the DM particle is also regarded
as captured. If one assumes that an equilibrium between capture and
reflection~(which includes evaporation in a more general sense, and
annihilation is negligible for a DM with mass $m_{\chi}<1\,\mathrm{GeV}$)
is reached today, the \textit{instant} reflection/evaporation velocity
spectrum can be obtained from the spectrum of a large number of simulated
reflection/evaporation events that occur over a finite time span.

In practice, it happens frequently that the sampled DM particle is
effectively trapped within the Sun in the optically thick parameter
regime, and thus a truncation on the simulated number of collisions
is necessary. In our practical computation, a cut-off is imposed on
the number of scattering $n_{\mathrm{sc}}=200$, which means if a
DM particle experiences more than $200$ scatterings it is considered
as captured, and the simulation is terminated. These captured DM particles
are supposed to be sufficiently thermalized, to evaporate subsequently
and hence also contribute to the reflection spectrum.

Since it is unpractical to further simulate the captured DM particles
until they are evaporated, we utilize the velocity spectrum of the
reflected ones in simulations to deduce that of the captured ones.
To be specific, we extrapolate from the statistics of reflection events
undergoing $200\geq n_{\mathrm{sc}}>50$ collisions, a number large
enough to assume a fully thermalization of an MeV DM particles, to
give a trustable description of the evaporation spectrum~(of the
captured ones), which is then combined with the reflection statistics
to give a total spectrum. To get a sense, in the left panel of Fig.~\ref{fig:VelocitySpectrum}
we present the probability density function~(PDF) of the scattering
numbers $n_{\mathrm{sc}}$ for an example DM particle with a mass
$m_{\chi}=1\,\mathrm{MeV}$, and a cross section $\sigma_{\chi e}=1\times10^{-32}\,\mathrm{cm}^{2}$.
It is evident that $n_{\mathrm{sc}}=200$ is a sufficiently large
cut-off in the sense that the majority of the reflection events can
be directly described from simulation. Even for those capture events,
the statistics of $n_{\mathrm{sc}}>50$ collisions can provide a reasonable
description of their evaporation spectrum.

In the optically thin regime~(i.e., $\sigma_{\chi e}<10^{-36}\,\mathrm{cm}^{2}$),
however, considering that only a negligible portion of DM particles~(in
the MeV mass range) are captured after more than a few scatterings,
in this case we only consider the reflection events, and neglect the
evaporation. 

\begin{figure*}
\begin{centering}
\includegraphics[scale=0.7]{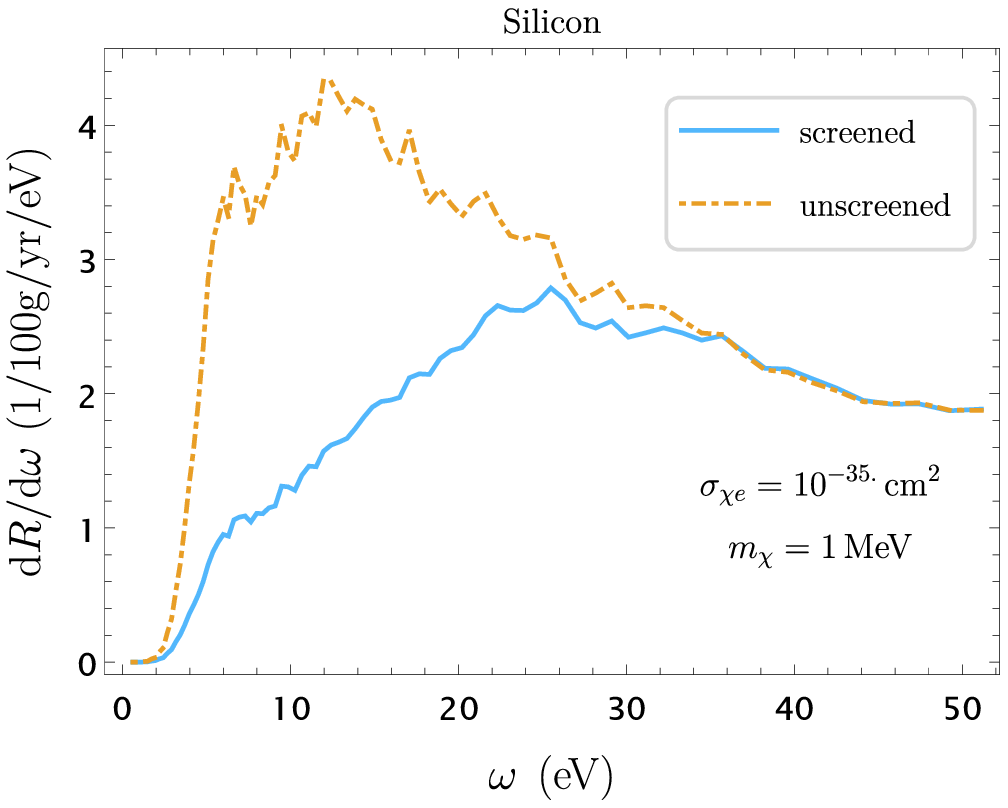}\hspace{1cm}\includegraphics[scale=0.73]{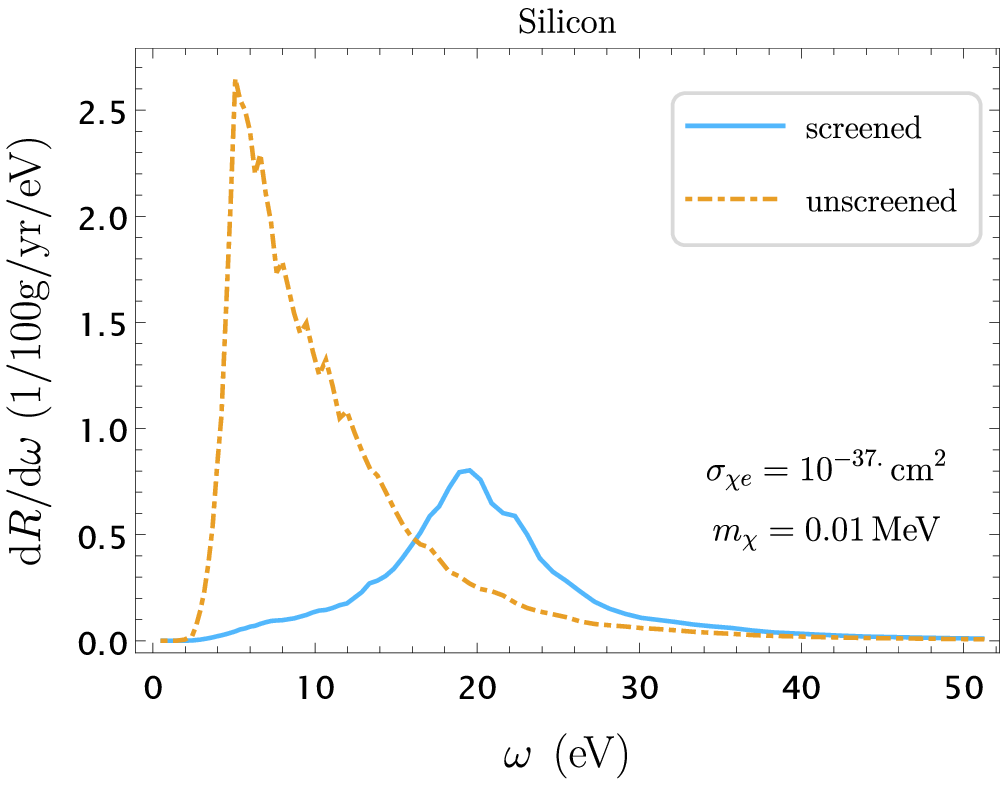}
\par\end{centering}
\caption{\label{fig:solar_spectrum}The differential excitation rate in silicon
target induced by the solar-reflected DM flux with~(\textit{blue})
and without~(\textit{orange dashed}) the screening effect, respectively,
for example DM with $m_{\chi}=1\,\mathrm{MeV}$ and cross section
$\sigma_{\chi e}=1\times10^{-35}\,\mathrm{cm}^{2}$~(\textbf{\textit{left}}),
and with $m_{\chi}=0.01\,\mathrm{MeV}$ and cross section $\sigma_{\chi e}=1\times10^{-37}\,\mathrm{cm}^{2}$~(\textbf{\textit{right}}),
respectively.}
\end{figure*}

Therefore, the differential flux of the solar reflected DM particle
can be expressed with simulation parameters as the following, 
\begin{eqnarray}
\frac{\mathrm{d}\Phi_{\mathrm{ref}}}{\mathrm{d}v}\left(v_{i}\right) & = & \frac{1}{4\pi D^{2}}\left(\frac{N_{i}}{N_{\mathrm{sample}}}\right)\frac{\Gamma\left(m_{\chi}\right)}{\Delta v_{i}},
\end{eqnarray}
where $N_{\mathrm{sample}}$ is the total number of the simulated
events, $N_{i}$ is event number collected in the $i$-th velocity
bin, with $v_{i}$ and $\Delta v_{i}$ being its center value and
its width, respectively, and $\Gamma\left(m_{\chi}\right)$ is obtained
through calculating the integral in Eq.~(\ref{eq:SurfaceRate}).
In order to formulate the experimental event rate from the solar reflection
in a parallel fashion to that of the galactic origin, it is necessary
to connect the differential reflection flux with the local density
of the reflected DM particles as follows,
\begin{eqnarray}
\frac{\mathrm{d}\Phi_{\mathrm{ref}}}{\mathrm{d}v}\left(v\right) & =n_{\oplus}vf_{\mathrm{ref}}\left(v\right) & ,\label{eq:Flux}
\end{eqnarray}
where $n_{\oplus}$ and $f_{\mathrm{ref}}\left(v\right)$ represent
the number density and the velocity distribution of the reflected
DM particles at the location of the Earth. Since the the Earth-Sun
distance is much larger than the radius of the Sun~($D\sim200\,R_{\odot}$),
the trajectories of the reflected DM particles can be approximated
as radially directed. Besides, the anisotropy effect of the halo DM
flux~(and hence the anisotropy of the reflected DM flux) is neglected
in above discussion. In the right panel of Fig.~\ref{fig:VelocitySpectrum}
we show the differential flux spectrum of a $1\,\mathrm{MeV}$ DM
particle, which, as a whole will appears in the formulation of experimental
excitation rate in the following discussions. It is understandable
that as the cross section turns smaller, DM particle has a higher
chance to reach the hotter core of the Sun, and thus be boosted to
a higher speed, as shown in Fig.~\ref{fig:VelocitySpectrum}.
\begin{spacing}{1.2}

\subsection{Screening effect in the detection of reflected DM particles}
\end{spacing}

The solar reflected DM particles can be probed with the terrestrial
detectors. Such detection strategy is especially preferred for the
DM particles in the MeV and sub-MeV mass range, where the DM particles
can effectively receive substantial kinetic energy from the hot solar
core, and hence are boosted over the conventional detector thresholds.
We first formulate the excitation rate of the solar reflection in
terms of the ELF, and quantitatively describe the screening effect
in relevant process.

By use of Eq.~(\ref{eq:Flux}), and substituting $\rho_{\chi}/m_{\chi}$
and $\mathrm{d}^{3}v\,f_{\chi}\left(\mathbf{v}\right)$ with $n_{\oplus}$
and $\mathrm{d}vf_{\mathrm{ref}}\left(v\right)$ respectively in Eq.~(\ref{eq:EventRateELF}),
it is straightforward to express the experimental event rate from
the solar reflection as follows, 
\begin{eqnarray}
R & = & n_{\oplus}\frac{\sigma_{\chi e}\,N_{\mathrm{cell}}}{4\,\alpha\,\mu_{\chi e}^{2}}\int\mathrm{d}\omega\int\mathrm{d}v\,\frac{f_{\mathrm{ref}}\left(v\right)}{v}\int\frac{\Omega\,\mathrm{d}^{3}Q}{\left(2\pi\right)^{3}}\,\nonumber \\
 &  & \times Q\,\mathrm{Im}\left[-\overline{\epsilon^{-1}}\left(Q,\omega\right)\right]\,\Theta\left[v-v_{\mathrm{min}}\left(Q,\,\omega\right)\right]\nonumber \\
 & = & \frac{\sigma_{\chi e}\,N_{\mathrm{cell}}}{4\,\alpha\,\mu_{\chi e}^{2}}\int\mathrm{d}\omega\int\frac{\mathrm{d}v}{v^{2}}\,\frac{\mathrm{d}\Phi_{\mathrm{ref}}\left(v\right)}{\mathrm{d}v}\int\frac{\Omega\,\mathrm{d}^{3}Q}{\left(2\pi\right)^{3}}\,\nonumber \\
 &  & \times Q\,\mathrm{Im}\left[-\overline{\epsilon^{-1}}\left(Q,\omega\right)\right]\,\Theta\left[v-v_{\mathrm{min}}\left(Q,\,\omega\right)\right].\label{eq:EvenRate_solar}
\end{eqnarray}

Now we can calculate the excitation rates of the solar reflection
in terms of the ELF. In the left~(right) panel in Fig.~\ref{fig:solar_spectrum}
shown are the differential rates in silicon target with an exposure
$100\,\mathrm{g}\cdot\mathrm{yr}$ for a benchmark DM mass $m_{\chi}=1\,\mathrm{MeV}$~($0.01\,\mathrm{MeV}$)
and a cross section $\sigma_{\chi e}=1\times10^{-35}\,\mathrm{cm}^{2}$~($1\times10^{-37}\,\mathrm{cm}^{2}$),
with and without the screening, respectively. In contrast to the case
of the halo DM where the event rates are significantly suppressed
in energy region $\omega>10\,\mathrm{eV}$~(see Fig.~\ref{fig:comparison}),
the spectra of the solar reflection extend to a higher energy range
beyond $50\,\mathrm{eV}$, a value corresponds to the ionization signal
$\mathcal{Q}=14e^{-}$. 

Intriguingly, as the velocity of reflected DM increases~(i.e., with
smaller cross section and smaller mass), the kinetically relevant
parameter space begins to overlay the plasmon pole, which features
a resonant peak~(blue curve) around $20\,\mathrm{eV}$ in the right
panel of Fig.~\ref{fig:solar_spectrum}. While such plasmon resonance
can not be directly produced in the scattering of the halo DM particles,
its observation has been expected in solar reflection and other possible
sources of fast DM particles~\citep{Kurinsky:2020dpb}. It is also
noted that unscreened differential rates still dominate in the low
energy region, and as a result the in-medium effects may remain suppressing
the total event rates when compared to the unscreened calculations.
\begin{figure}[h]
\centering{}\includegraphics[scale=0.58]{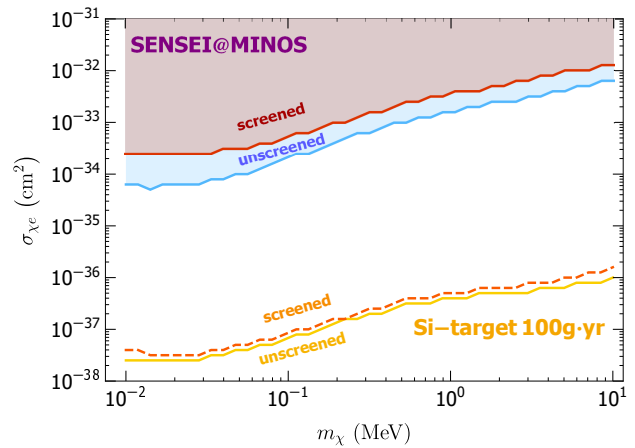}\caption{\label{fig:constraint}Exclusion curves~(90\%~C.L.) for DM-electron
cross section $\sigma_{\chi e}$ from the reflected DM flux. The filled
contours illustrate the constraints on $\sigma_{\chi e}$ from the
silicon-based SENSEI@MINOS experiment~\citep{Barak:2020fql} with~(\textit{red})
and without~(\textit{blue}) the screening, respectively. Also shown
are the 90\%~C.L. projected sensitivities of silicon semiconductor
with zero background and $100\,\mathrm{g\cdot yr}$ exposure with~(\textit{orange
dashed}) and without~(\textit{orange solid}) the screening, respectively.}
\end{figure}

Then, based on Eq.~(\ref{eq:EvenRate_solar}) and the released SENSEI@MINOS
results~\citep{Barak:2020fql}, which are presented as 90\% C.L.
limits on binned ionization signals $\mathcal{Q}=1e^{-},\,2e^{-},\,3e^{-},\,\mathrm{and}\,4e^{-}$,
respectively, we calculate the corresponding upper limits of the DM-electron
cross section $\sigma_{\chi e}$ in Fig.~\ref{fig:constraint}, in
both scenarios where the screening effect is neglected and accounted
for. Following the analysis in Ref.~\citep{Barak:2020fql}, parameters
$E_{g}=1.2\,\mathrm{eV}$, and $\varepsilon=3.8\,\mathrm{eV}$ are
adopted in deriving the SENSEI@MINOS constraints. The overall limits
are presented as the most stringent one from the four individual signals
bins. We also present the projections at 90\%~C.L. for a future silicon
detector with no background and an exposure of $100\,\mathrm{g}\cdot\mathrm{yr}$
in the signal window $\left[2e^{-},\,14e^{-}\right]$, for both the
screening and non-screening scenarios. The unscreened constraints
are found to be well consistent with those derived in Ref.~\citep{Emken:2021lgc}.
It is noted that the screening effect is less remarkable for the reflected
DM signals than for the Galactic signals, especially in the optically
thin regime, where the signals extend to even higher energies while
the screening takes effect in the low energy region. \vspace{0.5cm}

\section{\label{sec:conclusions}Summary and conclusions}

In this paper we perform a detailed derivation of the electronic excitation
event rate induced by the DM-electron interaction, taking into account
also the screening effect, which is described by the ELF, or the inverse
dielectric function. We take the EELS as an example to illustrate
how to generalize the discussion of a particle scattering problem
at zero temperature to the linear response theory description of the
target material exposed to bombardment by DM particles at a finite
temperature. In the latter framework the electronic many-body effects
are naturally encoded in the dielectric function. We then further
extend this procedure to formulate the material response to the DM
particles, and perform a DFT calculation for the diamond and silicon
targets.

Our numerical calculations not only verify the screening effect for
the two targets, but also depict the detailed dependence of screening
effect on the energy deposition $\omega$. To summarize, the screening
effect is remarkable in the low-energy regime, and as a result, the
prediction of excitation rates are suppressed by an $\mathcal{O}\left(1\right)$
factor compared to conventional approach in $\mathtt{QEDark}$~\citep{Essig:2015cda}.
In addition, we also explore the consequence of two different definitions
of the angular-averaged inverse dielectric function, namely, the ELFs
with and without LFEs. In the first case, one directly averages the
inverse of the dielectric matrix to obtain the inverse dielectric
function, while in the other case, one first averages the dielectric
matrix and then approximate its inverse as the inverse dielectric
function. A detailed calculation of diamond and silicon targets shows
that the differences between the excitation event rates estimated
from these two definitions are well within a factor of $0.5$, providing
a direct quantification of the LFEs.

Moreover, we compare the projected sensitivities for silicon calculated
using the $\mathtt{YAMBO}$ code with those obtained form the $\mathtt{GPAW}$
estimation~\citep{Knapen:2021run}. While in a broad range of DM
mass, the two approaches are found to be well consistent, a noticeable
discrepancy appears in the low-mass regime, which originates from
the operating parameters adopted in practical implementation. However,
such difference disappears if a $2e^{-}$ threshold is adopted in
experimental analysis.

In this study we also investigate the in-medium screening effect on
detecting the solar-boosted DM flux in silicon-based detectors, which
is a promising channel for the probe of MeV and sub-MeV DM particles.
With masses in this range, DM particles can be accelerated by the
energetic electrons in solar plasma to an energy in the keV scale,
so to be detected by conventional semiconductor detectors. Our calculations
show that the screening effect brings an $\mathcal{O}\left(0.1\sim1\right)$
reduction in excitation rates induced by the solar-boosted DM particles,
compared to the rates estimated by neglecting the screening. Besides,
our calculations verify the production of the plasmon resonance from
the scattering between semiconductor targets and fast solar-reflected
DM particles, especially in the optically thin regime.
\begin{acknowledgments}
This work was partly supported by Science Challenge Project under
No.~TZ2016001, by the National Key R\&D Program of China under Grant
under No.~2017YFB0701502, and by National Natural Science Foundation
of China under No.~11625415. C.M. was supported by the NSFC under
Grants No.~12005012, No.~11947202, and No.~U1930402, and by the
China Postdoctoral Science Foundation under Grants No.~2020T130047
and No.~2019M660016. 
\end{acknowledgments}

\appendix
\begin{spacing}{1.2}

\section{\label{sec:AppendixA}An estimate of DM-electron and DM-nucleus cross
sections in the Sun}
\end{spacing}

In this Appendix we give a detailed estimate of the DM-electron and
loop-induced DM-nucleus cross sections in the Sun, for both scalar~($S\otimes S$)
and vector~($V\otimes V$) interaction types discussed in Sec.~\ref{sec:Solar_Reflection}.
As mentioned in Sec.~\ref{sec:Solar_Reflection}, previous study~\citep{Kopp:2009et}
pointed out that even in a wide range of leptophilic DM models there
emerges loop-induced DM-hadron interactions, through photons emitted
from virtual leptons coupling to the charge of a nucleus. 

To summarize the result, first, the typical DM-electron cross section
$\sigma\left(\chi e\rightarrow\chi e\right)$ is expressed as $g_{\chi}^{2}g_{e}^{2}\times\sigma_{\chi e}^{0}$,
with
\begin{eqnarray}
\sigma_{\chi e}^{0} & = & \frac{\mu_{\chi e}^{2}}{\pi\Lambda^{4}},
\end{eqnarray}
and $\Lambda$ representing the energy scale of the heavy mediators\footnote{{\selectfont In Ref.~\citep{Kopp:2009et}, $\Lambda$ is set to
be $10\,\mathrm{GeV}$.}}, where perturbative couplings $g_{\chi}$ and $g_{e}$ are also defined.
For scalar~($S\otimes S$) type interaction, a DM-nucleus cross section
$\sigma\left(\chi N\rightarrow\chi N\right)$ appears at the two-loop
level, which is estimated as $\alpha^{2}g_{\chi}^{2}g_{e}^{2}\times\sigma_{\chi N}^{1}$~(see
Table~\uppercase\expandafter{\romannumeral1} in Ref.~\citep{Kopp:2009et}),
where the typical cross section $\sigma_{\chi N}^{1}$ can be written
in terms of $\sigma_{\chi e}^{0}$~(eq.~(26) in Ref.~\citep{Kopp:2009et})
\begin{eqnarray}
\sigma_{\chi N}^{1} & \approx & \left(\frac{\alpha\,Z}{\pi}\right)^{2}\left(\frac{\mu_{\chi N}}{\mu_{\chi e}}\right)^{2}\times\sigma_{\chi e}^{0},
\end{eqnarray}
with $\mu_{\chi N}=m_{N}\,m_{\chi}/\left(m_{N}+m_{\chi}\right)$ being
the reduced mass of the DM-nucleus pair; for vector~($V\otimes V$)
type interaction, the induced DM-nucleus cross section appears at
the one-loop level, which is estimated as $g_{\chi}^{2}g_{e}^{2}\times\sigma_{\chi N}^{1}$.
It is evident that in the WIMP direct detection, where $\mu_{\chi N}\sim m_{N}\sim m_{\chi}\sim\mathcal{O}\left(10\,\mathrm{GeV}\right)$,
thus $\sigma\left(\chi N\rightarrow\chi N\right)\gg\sigma\left(\chi e\rightarrow\chi e\right)$.
However, since in the Sun the DM particles predominantly scatter with
the hydrogen and helium nuclei~($Z\leq2$), which brings a suppression
factor $\sim10^{-2}$ compared to $\sigma_{\chi N}^{1}$, and more
importantly, a DM mass below $\mathcal{O}\left(10\,\mathrm{MeV}\right)$
significantly reduces the ratio $\left(\mu_{\chi N}/\mu_{\chi e}\right)^{2}$
to $\mathcal{O}\left(1\right)$. In addition, the ratio between the
thermal velocities of nuclei and solar electrons contributes another
factor $\sqrt{m_{e}/m_{N}}$ that further suppresses the DM-nucleus
scattering event rates. Thus, for the two benchmark DM models in this
study, it is safe to neglect the loop-induced DM-nucleus interaction.
\vspace{0.5cm}

\section{\label{sec:Appendix}Derivation of the rate $\Gamma$}

\begin{figure}
\begin{centering}
\includegraphics[scale=0.5]{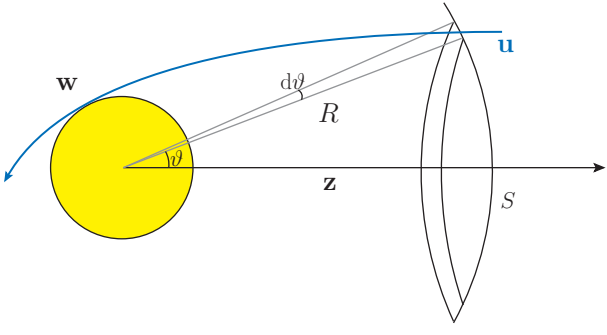}
\par\end{centering}
\caption{\label{fig:Sun_encounter}Illustration of a DM particle encounters
the Sun. The blue curve represents a trajectory of DM particle~(with
velocity $\mathbf{u}$ in the solar neighborhood) tangent to the solar
surface sphere, with a critical angular momentum $J_{c}=wR_{\odot}$.
It is evident that only those with angular momentum below $J_{c}$
can contribute to the encounter rate $\Gamma$. See text for details.}
\end{figure}

Here we provide a review on the derivation of the rate $\Gamma$ at
which the DM flux reaches the solar surface given in Eq.~(\ref{eq:SurfaceRate}).
As illustrated in Fig.~\ref{fig:Sun_encounter}, we imagine a swarm
of DM particles from phase space $f_{\chi}\left(\mathbf{u}\right)\mathrm{d}^{3}u$
enters the differential ring area opened by $\mathrm{d}\vartheta$,
at a rate of
\begin{eqnarray}
\mathrm{d}\Gamma & = & \frac{\rho_{\chi}}{m_{\chi}}\,uf_{\chi}\left(\mathbf{u}\right)\mathrm{d}^{3}u\times2\pi R^{2}\sin\vartheta\cos\vartheta\mathrm{d}\vartheta\nonumber \\
 & = & \frac{\rho_{\chi}}{m_{\chi}}\,\pi\frac{f_{\chi}\left(\mathbf{u}\right)\mathrm{d}^{3}u}{u}\times\mathrm{d}\left(uR\sin\vartheta\right)^{2}\nonumber \\
 & = & \frac{\rho_{\chi}}{m_{\chi}}\,\pi\frac{f_{\chi}\left(\mathbf{u}\right)\mathrm{d}^{3}u}{u}\times\mathrm{d}J^{2},\label{eq:differentialGamma}
\end{eqnarray}
where the $\mathbf{z}$-axis is by construction parallel to the velocity
$\mathbf{u}$, and radius $R$ of an imaginary sphere $S$ is so large
that the solar gravitational effect is negligible outside the sphere.
Of course, not all of the DM flux entering the ring end up reaching
the Sun, but Eq.~(\ref{eq:differentialGamma}) suggests that angular
momentum $J=uR\sin\vartheta$ is a convenient variable in analysis.

Now we calculate the rate at which above DM particle flux reaches
the solar surface. First it is straightforward to observe that for
those DM particles reach the solar surface with incident velocity
$\mathbf{u}$, their speed definitely equals $w=\sqrt{u^{2}+v_{\mathrm{esc}}^{2}\left(R_{\odot}\right)}$,
and thus there exists a maximum angular momentum $J_{c}=wR_{\odot}$.
That is to say, as the angle $\vartheta$ exceeds a certain critical
value, the trajectory of the DM particle will no longer intersect
with the solar surface, because a larger angle $\vartheta$ corresponds
to a larger angular momentum. This can be equivalently stated in terms
of angular momentum: only those DM particles with angular momentum
$J\leq wR_{\odot}$ have the chance to enter the Sun.

Although we are discussing the encounter rate of the DM flux with
a specific velocity $\mathbf{u}$, the last line in Eq.~(\ref{eq:differentialGamma})
turns out to be irrelevant of the illustrative $\mathbf{z}$-axis.
So for an arbitrary DM velocity $\mathbf{u}$, one can always construct
a specific $\mathbf{z}$-axis parallel to this $\mathbf{u}$, repeat
the above discussion, and will obtain the same form of differential
rate. Therefore, by first integrating over variable $J^{2}$ up to
$J_{c}^{2}$ in Eq.~(\ref{eq:differentialGamma}) and further integrating
over the DM velocity $\mathbf{u}$, we obtain the encounter rate in
Eq.~(\ref{eq:SurfaceRate}). Moreover, by conservation of the angular
momentum, we conclude that $J^{2}$ remains uniform, and the number
remains constant per $J^{2}$ interval for the DM particles entering
the sphere $S$ and entering the Sun, which helps us determine the
injection angle of the DM particles at the solar surface.

\twocolumngrid
\renewcommand{\baselinestretch}{1.1}

\bibliographystyle{JHEP1}
\addcontentsline{toc}{section}{\refname}\bibliography{SolarElectronScreening}

\end{document}